\documentclass[aps,pre,twocolumn,superscriptaddress,floats]{revtex4-2}

\usepackage{graphicx}
\usepackage{amsmath,amssymb}
\usepackage{hyperref}
\usepackage{overpic}

\usepackage{xcolor}
\usepackage{array}

\usepackage{tikz}
\usepackage[compat=1.1.0]{tikz-feynman}
\tikzfeynmanset{warn luatex=false}
\makeatletter
\def\tikzfeynman@luatex@required@path{}
\makeatother

\newcommand{\tif}{{\tilde { f}}}
\newcommand{\iu}{\mathrm{i}}
\newcommand{\eu}{\mathrm{e}}
\newcommand{\di}{\mathrm{d}}

\newcommand{\dder}[2]{\frac{\delta #1}{\delta #2}}

\newcommand{\bol}[1]{\boldsymbol{#1}}
\newcommand{\vx}{\bol{x}}
\newcommand{\vq}{\bol{q}}
\newcommand{\vk}{\bol{k}}

\newcommand{\tq}{\tilde {\bol{q}}}
\newcommand{\tp}{\tilde {\bol{p}}}
\newcommand{\tk}{\tilde {\bol{k}}}
\newcommand{\thh}{\tilde {\bol{h}}}
\newcommand{\vnabla}{\bol{\nabla}}

\newcommand{\hpsi}{\tilde{\psi}}      
\newcommand{\hrho}{\tilde{\rho}}      

\newcommand{\tD}{\tilde{D}}
\newcommand{\tlambda}{\tilde{\lambda}}
\newcommand{\tu}{\tilde{u}}
\newcommand{\Ham}{\mathcal{F}}

\begin{document}

\title{Non-equilibrium phase transition in the Brownian Ising Model:\\
       field theory, renormalization group, and exact results}

\author{Mattia Scandolo}
\affiliation{Laboratoire de Physique de l'École normale supérieure, ENS, Université PSL,
             CNRS, Sorbonne Université, Université de Paris, F-75005 Paris, France}
\affiliation{Department of Physics, University of Chicago, Chicago, Illinois 60637, USA}



\author{Luca Di Carlo}
\affiliation{Joseph Henry Laboratories of Physics, Princeton University,
             Princeton, New Jersey 08544, USA}
\affiliation{Lewis-Sigler Institute for Integrative Genomics, Princeton University,
             Princeton, New Jersey 08540, USA}

\date{\today}

\begin{abstract}
We present a complete field-theoretical renormalization-group (RG) analysis of the Brownian Ising Model (BIM), in which a $\mathbb{Z}_2$ order parameter is coupled to a passive conserved density, breaking detailed balance. Using the Martin--Siggia--Rose formalism and an $\epsilon=4-d$ expansion, we show that this density-order parameter coupling is RG-relevant below four dimensions and drives the system to a new non-equilibrium fixed point, distinct from the Ising universality class. Critical exponents are computed at lowest nontrivial order, some of which require a dedicated two-loop analysis. At large scales, the density acts as an effective noise that is white in time but long-range in space, enhancing order-parameter fluctuations and producing a negative anomalous dimension $\eta$. A defining feature of the new class is that the correlation and response functions acquire different anomalous dimensions, $\eta \neq 2 - \gamma/\nu$ — a direct, observable signature of fluctuation-dissipation-theorem violation at large scales that cannot occur in equilibrium. We also find a small correction-to-scaling exponent, implying large preasymptotic corrections that must be accounted for in numerical and experimental tests.
We further derive a set of relations among renormalization factors that hold to all orders in perturbation theory, following from the linearity of the density dynamics and an emergent shift symmetry. These yield an exact scaling relation $\nu = 2/(d+z-2)$ at the BIM fixed point and establish that the Ising universality class, as well as the diluted-Ising universality class, are unstable in $d=3$. This establishes the BIM fixed point as the unique infrared attractor for any nonzero diffusion constant.
\end{abstract}

\maketitle

\section{Introduction}\label{sec:intro}

The Ising universality class governs the critical behavior of equilibrium systems with a scalar order parameter and $\mathbb{Z}_2$ symmetry, from uniaxial ferromagnets to liquid-gas phase separation~\cite{fisher1967theory}. Its critical exponents in $d=3$ are among the best-determined quantities in theoretical physics, with values established to six-loop precision~\cite{kompaniets2017sixloop,pelissetto2002critical}.
A striking and less appreciated feature of the Ising universality class is its robustness to non-equilibrium perturbations. From the renormalization group (RG) perspective, the large-scale properties of the theory are controlled by a few structural features --- e.g. symmetry, dimensionality, conservation laws. Detailed balance represents one such structural feature, as it is related to the time-reversal symmetry of the steady-state measure. However, breaking detailed balance at the microscopic level does not necessarily break it at large scales: when non-equilibrium effects flow to zero under coarse-graining, the time-reversal symmetry is effectively restored~\cite{tauber2014criticaldynamics,tauber1997critical,bassler1994critical}. This has been verified in a wide range of systems: probabilistic cellular automata~\cite{grinstein1985pca}, kinetic spin systems driven by competing heat baths~\cite{odor2004review}, models with explicit nonreciprocal interactions \cite{dicarlo2025offequilibrium,garces2025phase,akritidis2026ising}, as well as active systems undergoing motility-induced phase separation~\cite{cates2015mips,caballero2018mips,maggi2022critical}. In all of these cases, the non-equilibrium perturbations are RG-irrelevant at the Ising fixed point~\cite{akkineni2004nonequilibrium}, and the large-scale universality class remains unchanged.
A well established route away from the equilibrium Ising universality class is quenched disorder: random impurities couple locally to the mass term of the order parameter and alter the universality class whenever the Harris criterion $d\nu > 2$ is violated~\cite{harris1974random}. Since the Ising exponent $\nu \approx 0.630 < 2/3$ in $d=3$~\cite{pelissetto2002critical}, the Ising fixed point is unstable to quenched impurities in three dimensions, giving rise to the diluted Ising universality class~\cite{khmelnitskii1975second,lubensky1975critical,grinstein1976disordered}. Crucially, however, the underlying mechanism is essentially equilibrium in character: the disorder is a static, externally imposed inhomogeneity.

In this paper we study in detail the \textit{Brownian Ising model} (BIM), a minimal field theory in which a $\mathbb{Z}_2$-symmetric order parameter $\psi$ is coupled to an independently evolving conserved density $\rho$. The defining structural feature is the complete absence of feedback from $\psi$ onto $\rho$: agents move independently of their internal state, so the density current carries no $\psi$-dependent terms. This nonreciprocal structure explicitly breaks detailed balance while exactly preserving the $\mathbb{Z}_2$ symmetry of $\psi$. As we demonstrate below, this structure introduces a qualitatively new route out of Ising universality that has no equilibrium analog.

The BIM is structurally distinct from Model~C of Hohenberg and Halperin~\cite{hohenberg1977theory}, which also couples a $\mathbb{Z}_2$ order parameter to a conserved scalar density but includes a magnetization-dependent density current $\bol{\nabla}\psi^2$; in the BIM that feedback is absent and the two theories belong to different universality classes (see Sec.~\ref{sec:modC} for a detailed comparison). The BIM describes a broad class of physical systems, most naturally bistable collections of motile particles in the weak-advection regime, as well as spatially extended bistable reaction networks embedded in strongly fluctuating environments where motile catalysts locally modulate reaction rates~\cite{solon2013flocking,solon2015flocking,chen2025bim, lima2022diffusivevoter}.

The critical behavior of the BIM is controlled by a competition between two hydrodynamic timescales. Near the critical point, the order parameter $\psi$ undergoes critical slowing down, with modes at wave-vector $q$ relaxing on a timescale $\tau_\psi \sim \lambda_0^{-1} \Lambda^{z-2} q^{-z}$. The conserved density is itself a slow hydrodynamic variable, relaxing diffusively as $\tau_\rho \sim D_0^{-1} q^{-2}$. Their ratio $\tau_\rho/\tau_\psi=(\lambda_{0}/D_{0})(q/\Lambda)^{z-2}$ controls the effective large-scale behavior of the model. Two distinct limits are of particular interest.
In the quenched limit $\tau_\rho/\tau_\psi\gg 1$, the density is effectively frozen on the timescales of $\psi$ and acts as quenched disorder — the theory reduces to Model A coupled to a random static field. In the opposite, fast-diffusion limit $\tau_\rho/\tau_\psi\ll 1$, density fluctuations relax far faster than $\psi$ and generate an effective noise with correlations $C^\mathrm{eff}_\rho(\vq, t)\sim q^{-2}\delta(t)$, which is white in time but carries long-range spatial correlations decaying as $r^{-(d-2)}$. Unlike short-range multiplicative noise — which is RG-irrelevant near $d_c = 4$ — the long-range character of $C_\rho^\mathrm{eff}$ makes it a relevant perturbation that destabilizes the Ising fixed point, driving the system to a new universality class. Our RG analysis shows that $z>2$, making the fast-diffusion limit the infrared (IR) fixed point of the BIM, and that the resulting universality class is distinct from both the Ising and quenched Ising classes.

A fundamental consequence of equilibrium is the fluctuation-dissipation theorem (FDT)~\cite{kubo1966fdt,tauber2014criticaldynamics}. At any equilibrium RG fixed point, the FDT constrains the structure of the fixed-point action such that the anomalous dimensions of the equal-time correlation function $C(\vq) \sim q^{-2+\eta}$ and the integrated response function $G(\vq) \sim q^{-2+\eta'}$ must coincide: $\eta = \eta'$, where $\eta'\equiv 2-\gamma/\nu$. A splitting $\eta \neq \eta'$ is therefore not merely a quantitative difference from Ising exponents — it is a definitive signature that non-equilibrium effects persist to arbitrarily large scales.
At the BIM fixed point we find $\eta = -\epsilon/8$ at one loop, while $\eta' = \mathcal{O}(\epsilon^2)$, so the splitting is already present at leading order in $\epsilon = 4-d$. Evaluating at $\epsilon = 1$ ($d=3$) gives $\eta \approx -0.125$ while $\eta' \approx 0.021$: a difference that could be directly measurable from the correlation and response functions of any microscopic realization of the BIM. The negative sign of $\eta$ is itself anomalous compared to the Ising class ($\eta > 0$): it reflects the fact that the long-range nature of $C_\rho^\mathrm{eff}$ \textit{enhances} correlations beyond the Gaussian level, rather than suppressing them.

The goal of this paper is to present in full detail the derivations of the following results that were summarized briefly in \cite{Scandolo2026BimShort}.

\textit{(i) New universality class.}
Using the Martin--Siggia--Rose path-integral formalism and a one-loop $\epsilon$-expansion~\cite{martin1973statistical,janssen1976on,de1976techniques, hohenberg1977theory}, we identify a new IR-stable RG fixed point: the BIM universality class. The critical exponents, including the FDT-violating splitting $\eta \neq \eta'$, are computed to lowest non-trivial order~(Sec.~\ref{sec:RG1}).

\textit{(ii) Exact relations.}
The linearity of the density equation implies that the density sector does not renormalize, yielding exact $Z$-factor identities valid to all orders in perturbation theory. Combined with an emergent Ward Identity of the effective field theory, these identities constrain the $\beta$-functions and lead to an exact scaling relation $\nu = 2/(d+z-2)$ at any perturbatively accessible fixed point with finite $\tif_R^*$~(Sec.~\ref{sec:exact}), even if $\epsilon$ is not small.

\textit{(iii) Generalized Harris criterion.}
The exact $\beta$-function for the coupling between $\psi$ and $\rho$ yields a stability criterion for the Ising fixed point: $(d+z-2)\nu > 2$ — a generalization of the Harris criterion to dynamical disorder, valid to all orders in perturbation theory. This criterion fully agrees with the findings of~\cite{vojta2016spatiotemporal}. Using this criterion together with known values of $z$ for Model A and the diluted Ising model in $d=3$, we rigorously rule out both the Ising and diluted-Ising fixed points as viable candidates for the large-scale behavior of the BIM in three dimensions~(Sec.~\ref{sec:exact}).
The same criterion predicts that the Ising fixed point is stable in two dimensions, consistent with existing numerical results~\cite{solon2015flocking,chen2025bim}.

\textit{(iv) Slow asymptotic convergence.}
The correction-to-scaling exponent $\omega \approx 0.020$ in $d=3$ is relatively small. Finite-size corrections to scaling observables decay as $L^{-\omega}$, and will thus be substantial even for large systems. We discuss the implications for numerical studies.

While our analysis is perturbative in $\epsilon = 4-d$, exact relations --- valid to all orders in perturbation theory --- allow us to draw conclusions about the stability of competing fixed points all the way down to $d=3$. A fully non-perturbative treatment, e.g., via the non-perturbative renormalization group (such as the functional renormalization group)~\cite{berges2002nprg,delamotte2007nprg}, would be required to rule out the emergence of additional non-perturbative fixed points at dimensions $2 < d < 4$, and is beyond the scope of the present work.

The remainder of the paper is organized as follows. In Sec.~\ref{sec:bim} we introduce the BIM, derive the hydrodynamic equations from first principles, and discuss the mean-field phase structure and the relation to Model C. Section~\ref{sec:FT} develops the field-theoretical formulation: the MSR action, the field rescaling to effective coupling constants, the Gaussian propagators, and the power-counting analysis that establishes $d_c = 4$. In Sec.~\ref{sec:RG1} we perform the complete one-loop renormalization, derive all anomalous dimensions and $\beta$-functions, identify the fixed points, and compute the critical exponents to lowest non-trivial order. Second order corrections are computed for exponents $z$, $\eta'$, and $\omega$, whose first-order corrections from mean field vanish. A special focus is also dedicated to the quenched-disorder limit, which reproduces the known results for the diluted Ising model. Section~\ref{sec:exact} presents the derivation of several results valid to all orders in perturbation theory: the $Z$-factor identities, the shift symmetry, and the generalized Harris criterion. We conclude in Sec.~\ref{sec:conclusion} with a discussion of open problems and broader implications.

\section{The Brownian Ising model}\label{sec:bim}

\subsection{Physical motivation and microscopic models}
The \textit{Brownian Ising model} (BIM) is a minimal field theory of a non-equilibrium system in which a $\mathbb{Z}_2$-symmetric order parameter $\psi$ is coupled to an independently evolving conserved scalar density $\rho$. Its defining structural feature is the complete absence of feedback from $\psi$ onto $\rho$: agents whose density is tracked diffuse \textit{independently} of their internal state, which is encoded in $\psi$. This nonreciprocal coupling explicitly breaks detailed balance while leaving the $\mathbb{Z}_2$ symmetry of $\psi$ intact.

The most natural class of BIM realizations consists of bistable collections of motile particles that diffuse in space while interacting locally with their neighbors. Each agent carries an internal degree of freedom — a local spin — that can flip between two states. Crucially, the spatial motion of the agents is governed entirely by isotropic diffusion, with no drift nor bias that depends on the spin state. Coarse-graining over many agents yields a local density field $\rho(\vx,t)$ that evolves purely diffusively, and a local order parameter $\psi(\vx,t)$ whose relaxation is modulated by the local density. The active Ising model~\cite{solon2013flocking,solon2015flocking} in the limit of weak advection — where particle transport is dominated by diffusion rather than alignment-induced drift — provides a concrete lattice realization of this phenomenology. Exact hydrodynamic equations for such systems, derived using the Doi-Peliti path-integral formalism~\cite{peliti1985path,kourbane2018exact,scandolo2023active}, reduce to the BIM equations in the diffusion-dominated limit. Related systems with BIM structure include diffusive voter models and active lattice gases in the weak-advection regime~\cite{lima2022diffusivevoter,chen2025bim}.

A second class of realizations arises in spatially extended bistable reaction networks embedded in a strongly fluctuating environment. The fluctuating environment can be represented, for example, by a density of motile catalysts, which locally modulate the reaction rates and hence the effective distance to the ordering transition. If the catalysts diffuse independently of the reaction products — as is natural when they are not consumed by the reaction — the resulting hydrodynamics again takes the BIM form. In this context, $\psi$ is the Ising-like order parameter representing the local concentration difference between the two stable phases, while $\rho$ represents the local concentration of catalysts. Higher concentrations of catalysts reduce the energetic barrier for jumping between the two stable phases. At a coarse-grained level, this corresponds to a dependence of the field-theoretical mass term $r$ on the local catalyst density $\rho$.

\subsection{Hydrodynamic equations}\label{sec:hydro}

\subsubsection{The conserved density}
The conserved density $\rho(\vx,t)$ evolves autonomously, without any feedback from the ordering field $\psi$. In the absence of external potentials or alignment forces, the relevant dynamics is purely diffusive. The continuity equation reads
\begin{align}
    \partial_t \rho &= -\vnabla\cdot\bol{J}\,, &
    \bol{J} &= -D_0\,\vnabla\rho + \sqrt{2D_0\rho}\,\bol{\zeta}\,,
    \label{eq:rho_full}
\end{align}
where $D_0$ is the bare diffusion constant and $\bol{\zeta}(\vx,t)$ is a Gaussian white noise, delta-correlated in space and time, $\langle\zeta_i(\vx,t)\zeta_j(\vx',t')\rangle = \delta_{ij}\delta^d(\vx-\vx')\delta(t-t')$. The noise amplitude $\sqrt{2D_0\rho}$ is set by the fluctuation-dissipation theorem for the diffusing particles, and encodes the Poissonian statistics of the underlying particle process.

When the mean density $\rho_0 > 0$, decomposing $\rho = \rho_0 + \delta\rho$ and linearizing in $\delta\rho$ yields the equation for density fluctuations,
\begin{align}
    \partial_t \delta\rho &= -\vnabla\cdot\bol{J}'\,, &
    \bol{J}' &= -D_0\,\vnabla\delta\rho + \sqrt{2\tD_0}\,\bol{\zeta}\,,
    \label{eq:deltarho}
\end{align}
where $\tD_0 \equiv D_0\rho_0$. The multiplicative noise correction $\sqrt{2D_0\delta\rho}$ present in Eq.~\eqref{eq:rho_full} is RG-irrelevant near $d_c=4$ and is discarded~\cite{scandolo2023active}. Equation~\eqref{eq:deltarho} is linear in $\delta\rho$ and can therefore be solved exactly. In Fourier space,
\begin{equation}
    \delta\rho(\vq,\omega) =
    \frac{\sqrt{2\tD_0}\;\iu\,\vq\cdot\bol{\zeta}(\vq,\omega)}{-\iu\omega + D_0 q^2}\,,
    \label{eq:rho_fourier}
\end{equation}
from which the density--density correlator follows immediately:
\begin{equation}
    \begin{split}
        \langle\delta\rho(\vq,t)\,\delta\rho(\vk,0)\rangle &= (2\pi)^d\delta^d(\vq+\vk)\,C_\rho(\vq,t)\,,\\
         C_\rho(\vq,t) &= \rho_0\,\eu^{-D_0 q^2|t|}\,.
    \end{split}
    \label{eq:Crho}
\end{equation}
The equal-time amplitude of $C_{\rho}(q)$ is set by the mean density $\rho_0 = \tD_0/D_0$.

\subsubsection{The order parameter and the RG-relevance of the mass coupling}
In the absence of density fluctuations (uniform and static $\rho = \rho_0$), the order parameter $\psi$ of a bistable system is expected to relax according to the standard Model~A dynamics~\cite{hohenberg1977theory}. Performing a Landau expansion in $\psi$ and $\vnabla\psi$ near the ordering transition yields
\begin{equation}
    \lambda_0^{-1}\partial_t\psi = \vnabla^2\psi - r_0\psi - \tu_0\psi^3
    + \sqrt{2\tlambda_0\lambda_0^{-2}}\,\xi\,,
    \label{eq:psi_ModelA}
\end{equation}
where $\xi(\vx,t)$ is a unit Gaussian white noise, $\lambda_0$ is the bare kinetic coefficient (inverse relaxation rate), $\tlambda_0$ sets the noise amplitude, $r_0$ is the bare mass (distance to criticality), and $\tu_0 > 0$ is the quartic coupling.

When $\rho$ is non-uniform, all kinetic coefficients in Eq.~\eqref{eq:psi_ModelA} are in principle functions of the local density. Expanding each coefficient around $\rho_0$,
\begin{equation}
    c(\rho_0 + \delta\rho) = c_0 + c'(\rho_{0})\delta\rho + \mathcal{O}(\delta\rho^2)\,,
\end{equation}
generates a set of interaction terms between $\psi$ and $\delta\rho$. The RG-relevance of each such term near $d_c=4$ is determined by its canonical scaling dimension. A standard power-counting analysis of the resulting field theory (carried out in Sec.~\ref{sec:FT}) shows that the only \textit{marginal} coupling at upper critical dimension $d=d_{c}$ is the one arising from the density dependence of the mass term:
\begin{equation}
    g_0 \equiv \left.\frac{d r}{d\rho}\right|_{\rho_0}\,.
    \label{eq:g0def}
\end{equation}
The corrections arising from the $\delta\rho$ dependence of $\lambda$, $\tu$, and $\tlambda$ each carry additional powers of frequency or momentum at the Gaussian fixed point and are therefore RG-irrelevant~\cite{scandolo2023active}. Dropping these irrelevant terms, the equation of motion for $\psi$ becomes
\begin{align}
    \lambda_0^{-1}\partial_t\psi &= \vnabla^2\psi - r_0\psi - \tu_0\psi^3
    - g_0\psi\,\delta\rho
    + \sqrt{\frac{2\tlambda_0}{\lambda_0^{2}}}\,\xi\,.
    \label{eq:psi}
\end{align}
Here the subscript $0$ on the couplings denotes that all parameters are evaluated at the mean density $\rho_0$, and hence do not carry any hidden $\rho$ dependence. Together with Eq.~\eqref{eq:deltarho}, this constitutes the hydrodynamic equations of the BIM. We note that Eqs.~\eqref{eq:deltarho} and~\eqref{eq:psi} agree with exact hydrodynamic equations derived for specific active Ising models in the small-advection limit~\cite{kourbane2018exact,scandolo2023active}.

The coupling $g_0\psi\,\delta\rho$ has a transparent physical interpretation: local density fluctuations shift the effective mass (distance to criticality) experienced by $\psi$. Since in thermal systems $r_0 \propto (T - T_0)/T_0$, with $T_{0}$ being the mean-field critical temperature, the coupling $g_0\psi\delta\rho$ effectively subjects $\psi$ to a spatiotemporally fluctuating temperature. This is structurally analogous to the field theory of the diluted Ising model~\cite{khmelnitskii1975second,lubensky1975critical,grinstein1976disordered,grinstein1977dynamics}, in which quenched impurities also enter through the mass term. The crucial difference is that in the BIM the disorder is \textit{dynamical} — the density field evolves in time — rather than static.

\subsection{Mean-field phase structure}\label{sec:mf}
The mean-field phase structure of the BIM follows from the spatially homogeneous, noiseless solutions of Eqs.~\eqref{eq:deltarho}--\eqref{eq:psi}. Setting $\delta\rho = 0$ and $\vnabla = 0$, the equation for the spatially averaged magnetization reduces to
\begin{equation}
    \lambda_0^{-1}\partial_t \psi_0 = -r_0\psi_0 - \tu_0\psi_0^3\,,
\end{equation}
which has the form of a standard pitchfork bifurcation. For $r_0 > 0$, the disordered state $\psi_0 = 0$ is the unique stable state. For $r_0 < 0$, two symmetric ordered states $\psi_0 = \pm\sqrt{-r_0/\tu_0}$ emerge and the disordered state becomes unstable. Fluctuations shift the mean-field critical point $r_0=0$ to the true critical value $r_c < 0$; we define
\begin{equation}
    \tau_0 \equiv r_0 - r_c
    \label{eq:tau0}
\end{equation}
as the control parameter measuring the true distance from criticality.

A linear stability analysis of the homogeneous ordered state against spatially inhomogeneous perturbations confirms that the ordered phase is stable: unlike the active Ising model~\cite{solon2015flocking}, where a traveling-band instability preempts the continuous transition near the ordering threshold, the BIM exhibits no such instability. The ordering transition is therefore continuous, and the ordered phase is spatially uniform at mean-field level.

\subsection{Timescale separation and the effective noise}\label{sec:timescale}
As anticipated in the introduction, the critical behavior of Eqs.~\eqref{eq:deltarho}--\eqref{eq:psi} is controlled by the competition between two hydrodynamic timescales: the order parameter relaxation time $\tau_\psi$ and the density relaxation time $\tau_\rho$. The ratio of these timescales introduces a dimensionless parameter, $w_{0}$, that will play a central role in the RG analysis.

\subsubsection{The timescale ratio}
Near the critical point, both $\psi$ and $\delta\rho$ are hydrodynamic slow variables whose characteristic frequencies vanish as $q\to0$. The order parameter $\psi$ undergoes critical slowing down, with modes at wave-vector $q$ relaxing on a timescale
\begin{equation}
    \tau_\psi \sim \lambda_0^{-1}\Lambda^{z-2}\,q^{-z}\,,
\end{equation}
where $\Lambda$ is the UV momentum scale (e.g., inverse lattice spacing) and $z$ is the dynamic critical exponent. Density modes, on the other hand, relax diffusively
\begin{equation}
    \tau_\rho \sim D_0^{-1}\,q^{-2}\,.
\end{equation}
The timescale ratio,
\begin{equation}
    \frac{\tau_\rho}{\tau_\psi} = w_0\left(\frac{q}{\Lambda}\right)^{z-2}\,,
    \qquad w_0 \equiv \frac{\lambda_0}{D_0}\,,
    \label{eq:w0}
\end{equation}
controls the relative dynamics of the two fields. At microscopic scales $q\sim\Lambda$, $w_0$ is the bare ratio of relaxation rates. In the hydrodynamic limit $q\to0$, the time-scale competition is entirely controlled by $z$, as the asymptotic behavior of $\tau_\rho/\tau_\psi$ depends on the sign of $z-2$. If $z<2$, density modes relax faster than order parameter modes at long wavelengths, while if $z>2$ the opposite is true.  Strong dynamic scaling, in which both fields relax on the same timescale, requires $z=2$. Which of these regimes is realized at large scales is therefore decided by the fixed-point value of $z$, computed in Sec.~\ref{sec:RG1}. We find $z>2$, indicating the fast-diffusion limit controls the large-scale dynamics. We examine the two limiting regimes — quenched ($\tau_\rho/\tau_\psi\gg1$) and fast-diffusion ($\tau_\rho/\tau_\psi \ll 1$) — in what follows.

\subsubsection{The quenched-disorder limit}\label{sec:quenched_limit}
Whenever $w_0(q/\Lambda)^{z-2}\gg1$, the density modes relax on timescales much longer than $\tau_\psi$. This would be the case, for example, if $z<2$ and $q$ is sufficiently small compared to $\Lambda$, or if $w_0\to\infty$. From the perspective of the order parameter, in this regime the density is effectively frozen. In other words, when observed on timescales $t\sim\tau_\psi$, Eq.~\eqref{eq:Crho} reduces to,
\begin{equation}
    C_\rho(\vq,t) \approx \rho_0\,,
    \label{eq:Crho_quenched}
\end{equation}
describing spatially uncorrelated, time-independent (quenched) disorder. The theory then reduces to Model~A with a random quenched mass — the universality class of the diluted Ising model~\cite{lubensky1975critical,grinstein1976disordered}. 

\subsubsection{The fast-diffusion limit and the effective noise}
\label{sec:fast_diff}
In the opposite regime, $w_0(q/\Lambda)^{z-2}\ll1$, density fluctuations relax much faster than the order parameter. This would be the case, for example, if $z>2$ and $q$ is sufficiently small compared to $\Lambda$, or if $w_0\to0$. Setting $w_0=0$ in Eq.~\eqref{eq:Crho} drives $C_\rho\to0$, suggesting that $\delta\rho$ decouples from $\psi$ whenever $w_0(q/\Lambda)^{z-2}\ll1$.
Although this is consistent with the naive expectation that fast modes decouple from slow dynamics, our RG analysis shows this expectation to be incorrect.


To understand why, consider the following argument. Over a time window $\delta t$, satisfying $\tau_\rho \ll \delta t \ll \tau_\psi$, the order parameter barely changes while density fluctuations fully decorrelate: $C_\rho(\vq,t)\approx 0$ for $t\gtrsim\delta t$. However, within this time window (infinitesimal from the point of view of $\psi$), the order parameter feels the \textit{integrated} effect of the fast density fluctuations. The effective contribution to the $\psi$ dynamics is thus determined by the time-integrated correlator,
\begin{equation}
    C^\mathrm{eff}_\rho(\vq,t\lesssim\delta t) =
    \frac{1}{\delta t}\int_0^{\delta t}\!\di t'\,C_\rho(\vq,t')
    \approx \frac{2\tD_0}{D_0^2 q^2\,\delta t}\,,
\end{equation}
with the last approximation following from the fact that $\delta t \gg \tau_\rho$. Taking the limit $\delta t/\tau_\psi\to0$ (i.e., $\delta t$ small on the scale of $\tau_\psi$) yields the effective correlator
\begin{equation}
    C^\mathrm{eff}_\rho(\vq,t) = \frac{2\tD_0}{D_0^2\,q^2}\,\delta(t)\,.
    \label{eq:effnoise}
\end{equation}
This result can also be obtained directly from Eq.~\eqref{eq:deltarho} by adiabatically eliminating the time derivative $\partial_t\delta\rho$, which is justified when $\delta\rho$ relaxes on timescales much shorter than $\tau_\psi$. The adiabatic elimination yields
\begin{equation}
    \delta\rho(\vq,t) = \sqrt{2\tD_0}\,\iu\,\vq\cdot\bol{\zeta}(\vq,t)/(D_{0}q^2)\,,
\end{equation}
which has the correlator given by Eq.~\eqref{eq:effnoise}.

At fixed $\lambda_{0}$, $C^\mathrm{eff}_\rho(\vq,t)$ vanishes in the $w_0\to0$ limit only if $\tD_{0}/D_{0}$ is held constant. As we shall discuss later in the RG analysis (Secs.~\ref{sec:RG1} and~\ref{sec:exact}), the large-scale behavior of the BIM is described by an effective theory with vanishing (renormalized) $w_{R}=0$, but a diverging ratio $\tD/D$ that precisely compensates the vanishing of $w$ to yield a finite effective noise amplitude. In other words, in the BIM the effective fast-diffusion limit at large scales is not a decoupling limit. The effective density correlator~\eqref{eq:effnoise} is white in time (fast) but carries \textit{long-range spatial correlations}, decaying in real space as $r^{-(d-2)}$ ($q^{-2}$ in Fourier space). This is qualitatively different from a short-range multiplicative noise: while short-range multiplicative noise is RG-irrelevant near $d_c=4$, the $q^{-2}$ spectrum of $C^\mathrm{eff}_\rho$ is RG-relevant below $d=4$ and destabilizes the Ising fixed point. The long-range nature of $C^\mathrm{eff}_\rho$ also explains the exotic result $\eta < 0$ at the BIM fixed point. The negative anomalous dimension reflects the enhanced large-scale correlations induced by this long-range noise — correlations of $\psi$ decay \textit{more slowly} than in the Gaussian theory.

\subsection{Relation to Model C}\label{sec:modC}
The BIM is closely related to, but structurally distinct from, Model~C of Hohenberg and Halperin~\cite{hohenberg1977theory}, which also describes a $\mathbb{Z}_2$ order parameter coupled to a conserved scalar density. In Model~C, the density current contains a magnetization-dependent term $\propto\vnabla\psi^2$, coupling the spatial dynamics of the density to the ordering field. In the BIM this term is absent, because agents diffuse independently of their internal state.

Importantly, this distinction is preserved under coarse-graining. No $\psi$-dependent contribution to the density current can be generated by the RG if it is absent microscopically. Formally, this follows from the field-theoretical observation that no diagram contributing to the vertex function $\Gamma_{\hrho \psi^2}$ can be generated. Consequently, the BIM and Model~C belong to distinct universality classes.

One can write the most general RG-relevant field theory coupling a $\mathbb{Z}_2$ order parameter to a conserved scalar density, compatible with isotropy and $\psi\to-\psi$:
\begin{align}
    \lambda_0^{-1}\partial_t\psi &= \vnabla^2\psi - r_0\psi - \tu_0\psi^3
    - g_0\psi\,\delta\rho + \sqrt{\frac{2\tlambda_0}{\lambda_0^2}}\,\xi\,,
    \label{eq:gen_psi}\\
    D_0^{-1}\partial_t\delta\rho &= \vnabla^2\delta\rho
    + \frac{g_0'}{2}\vnabla^2\psi^2
    + \sqrt{\frac{2\tD_0}{D_0^2}}\,\vnabla\cdot\bol{\zeta}\,.
    \label{eq:gen_rho}
\end{align}
Setting $g_0'=0$ recovers the BIM; general $g_0'\neq0$ gives the theory that interpolates between the BIM and Model~C.

Detailed balance is restored — and the theory becomes an equilibrium Model~C — when the noise amplitudes and coupling constants satisfy~\cite{akkineni2004nonequilibrium}
\begin{equation}
    \Theta \equiv \frac{\tlambda_0 D_0}{\tD_0 \lambda_0}\,\frac{g_0'}{g_0} = 1\,.
    \label{eq:Theta}
\end{equation}
In this case, an appropriate rescaling of the amplitudes of $\psi$ and $\delta\rho$ maps the theory to the purely relaxational dynamics:
\begin{align}
    \lambda_0^{-1}\partial_t\psi &= -\dder{\Ham}{\psi}
    + \sqrt{2\lambda_0^{-1}}\,\xi\,,\\
    D_0^{-1}\partial_t\delta\rho &= \vnabla^2\dder{\Ham}{\delta\rho}
    + \sqrt{2D_0^{-1}}\,\vnabla\cdot\bol{\zeta}\,,
\end{align}
with respect to the free-energy functional
\begin{equation*}
    \Ham = \int\!\di^d x\!\left[\frac{(\vnabla\psi)^2}{2} + \frac{r_0}{2}\psi^2
    + \frac{\tu_0}{4}\psi^4 + \frac{(\delta\rho)^2}{2}
    + g_0\,\psi^2\delta\rho\right]\!.
    \label{eq:freeenergy}
\end{equation*}
Physical realizations of this equilibrium theory include ferromagnetic spin fluids with Lennard-Jones interactions~\cite{nijmeijer1998isingfluid,nijmeijer1998heisenbergfluid} and active systems in which detailed balance is restored in the passive limit~\cite{agranov2024thermodynamically}.

Weak violations of detailed balance ($\Theta$ close to but not equal to $1$) have been shown to be RG-irrelevant~\cite{akkineni2004nonequilibrium}: the parameter $\Theta$ flows to its equilibrium value, $\Theta =1$, under the RG and the large-scale behavior remains in the Model~C universality class. The BIM corresponds to the opposite extreme $\Theta = 0$ (i.e., $g_0' = 0$), representing a \textit{strong} and structurally robust violation of detailed balance. As demonstrated below, this strong violation is RG-relevant and drives the system to a genuinely non-equilibrium universality class distinct from both Ising and Model~C.

\section{Field-theoretical formulation}\label{sec:FT}

\subsection{The Martin-Siggia-Rose action}
We cast the stochastic equations~\eqref{eq:deltarho} and~\eqref{eq:psi} into a field-theoretical path integral using the Martin--Siggia--Rose (MSR) formalism~\cite{martin1973statistical,janssen1976on,de1976techniques}. This introduces auxiliary response fields $\hpsi(\vx,t)$ and $\hrho(\vx,t)$ conjugate to $\psi$ and $\delta\rho$, respectively. Physical observables and correlation functions are then obtained as path-integral averages with statistical weight $\eu^{-S}$, where the action $S = \int\!\di^d x\,\di t\,\mathcal{L}$ decomposes as $\mathcal{L} = \mathcal{L}_0 + \mathcal{L}_I$. The Gaussian part, encoding the linear terms of Eqs.~\eqref{eq:deltarho} and~\eqref{eq:psi}, reads
\begin{equation}
\begin{split}
    \mathcal{L}_0 =\;
    &\hpsi\!\left(\lambda_0^{-1}\partial_t - \vnabla^2 + r_0\right)\!\psi
    - \tlambda_0\lambda_0^{-2}\,\hpsi^2\\
    +\;&\hrho\!\left(D_0^{-1}\partial_t - \vnabla^2\right)\!\delta\rho
    - \tD_0 D_0^{-2}\,(\vnabla\hrho)^2\,,
    \label{eq:L0}
\end{split}
\end{equation}
while the interaction part, encoding the nonlinear couplings, is
\begin{equation}
    \mathcal{L}_I = \tu_0\,\hpsi\,\psi^3 + g_0\,\hpsi\,\psi\,\delta\rho\,.
    \label{eq:LI}
\end{equation}
The $\hpsi^2$ term in $\mathcal{L}_0$ arises from the noise in the $\psi$ equation; the $(\vnabla\hrho)^2$ term encodes the (conserved) noise in the $\delta\rho$ equation. Causality is enforced by the standard MSR construction: $\langle\hpsi\,\hpsi\rangle_0 = \langle\hrho\,\hrho\rangle_0 = 0$ at the Gaussian level.

\subsection{Field rescaling and effective coupling constants}
\label{sec:rescaling}
The action just written contains seven independent bare parameters: $\lambda_0$, $\tlambda_0$, $D_0$, $\tD_0$, $r_0$, $\tu_0$, and $g_0$. However, only four independent combinations appear in the perturbative expansion, i.e. in the Feynman diagrams. To make this explicit, we rescale the fields by
\begin{align}
    \psi   &\to \sqrt{\frac{\lambda_0}{\tlambda_0}}\;\psi\,, \quad&
    \hpsi  &\to \sqrt{\frac{\tlambda_0}{\lambda_0}}\;\hpsi\,,
    \label{eq:rescale_psi}\\
    \delta\rho &\to g_0^{-1}\,\delta\rho\,, \quad&
    \hrho  &\to g_0\,\hrho\,.
    \label{eq:rescale_rho}
\end{align}
After this rescaling, the action takes the form $\mathcal{L} = \mathcal{L}_0' + \mathcal{L}_I'$ with
\begin{equation}
\begin{split}
    \mathcal{L}_0' =\;
    &\hpsi\!\left(\lambda_0^{-1}\partial_t - \vnabla^2 + r_0\right)\!\psi
    - \lambda_0^{-1}\,\hpsi^2\\
    +\;&\hrho\!\left(\lambda_0^{-1}w_0\,\partial_t - \vnabla^2\right)\!\delta\rho
    - \lambda_0^{-1}w_0 f_0\,(\vnabla\hrho)^2\,,
    \label{eq:L0_rescaled}
\end{split}
\end{equation}
\begin{equation}
    \mathcal{L}_I' = u_0\,\hpsi\,\psi^3 + \hpsi\,\psi\,\delta\rho\,,
    \label{eq:LI_rescaled}
\end{equation}
where the four effective bare coupling constants are
\begin{align}
    u_0 &= \frac{\tlambda_0}{\lambda_0}\,\tu_0\,, &
    f_0 &= \frac{\tD_0}{D_0}\,g_0^2\,, &
    w_0 &= \frac{\lambda_0}{D_0}\,, &
    r_0\,.
    \label{eq:effective_bare}
\end{align}
While $\lambda_0$ is still present in $\mathcal{L}_0'$, a rescaling of the frequency integration variable shows that it does not appear in any $Z$ factor; it can only contribute as an overall prefactor to the $\Gamma$ functions (see Appendix~\ref{app:oneloop}). All perturbative corrections therefore depend on $u_0$, $f_0$, $w_0$, and $r_0$ only.

The physical meaning of these effective parameters is as follows. The coupling $u_0$ is the standard quartic self-coupling of the Landau theory. The coupling $f_0 = \rho_0 g_0^2$ (with $\rho_0 = \tD_0/D_0$) quantifies the effective strength of density fluctuations as seen by $\psi$: it combines the amplitude of density fluctuations $\rho_0$ with the square of the coupling $g_0$. Neither $g_0$ alone nor $\rho_0$ alone determines the strength of the coupling; only their product $f_0$ is physically meaningful. The parameter $w_0$ is the bare timescale ratio introduced in Eq.~\eqref{eq:w0}, and $r_0$ is the bare mass.

\subsection{Gaussian propagators}\label{sec:propagators}
The Gaussian theory $\mathcal{L}_0'$, Eq.~\eqref{eq:L0_rescaled}, is exactly solvable. The statistical properties of the non-interacting system are entirely encoded in the 2-point functions. The only non-vanishing 2-point functions at the Gaussian level are the response functions $\langle\hpsi\,\psi\rangle_0$ and $\langle\hrho\,\delta\rho\rangle_0$, as well as the correlation functions $\langle\psi\,\psi\rangle_0$ and $\langle\delta\rho\,\delta\rho\rangle_0$. Correlations involving the response fields only, such as $\langle\hpsi\,\hpsi\rangle_0$ and $\langle\hrho\,\hrho\rangle_0$, are identically zero due to causality. All other 2-point functions vanish at the Gaussian level as well.

In Fourier space, with the convention $\tilde q = (q,\omega_{q})$, the response and correlation functions can be written as
\begin{equation}
    \langle\tilde\phi(\tq)\,\phi(\tk)\rangle_0\, = \tilde\delta(\tq+\tk) G_{\phi}^0(\tq)\,,
\end{equation}
\begin{equation}
    \langle\phi(\tq)\,\phi(\tk)\rangle_0\, =  \tilde\delta(\tq+\tk) C_{\phi}^0(\tq)\,,
\end{equation}
for both $\phi=\psi,\rho$. Here $\tilde\delta(\tq)\equiv (2\pi)^{d+1} \delta^{(d)}(q)\delta(\omega_{q})$.

For the order-parameter sector, the retarded propagator and the noise correlator are
\begin{align}
    G_\psi^0(q,\omega) &= \frac{1}{-\iu\lambda_0^{-1}\omega + q^2 + \tau_0}\,,
    \label{eq:Gpsi}\\
    C_\psi^0(q,\omega) &= 2\lambda_0^{-1}\,\bigl|G_\psi^0(q,\omega)\bigr|^2\,.
    \label{eq:Cpsi}
\end{align}
The function $G_\psi^0(q,\omega)$ is the retarded linear response of $\psi$ to a small external field~\cite{martin1973statistical,janssen1976on,de1976techniques}. The function $C_\psi^0(q,\omega) = \langle\psi\,\psi\rangle_0$ is the two-point correlator of the stochastic field, encoding the power spectrum of fluctuations. The relation $C_\psi^0(q,\omega) = 2\lambda_0^{-1}|G_\psi^0(q,\omega)|^2$ is a statement of the fluctuation-dissipation theorem at the Gaussian level; as we shall see, this relation is broken at the BIM fixed point and does not hold for the renormalized functions $G_\psi$ and $C_\psi$.

For the density sector, the analogous expressions are
\begin{align}
    G_\rho^0(q,\omega) &= \frac{1}{-\iu\lambda_0^{-1}w_0\,\omega + q^2}\,,
    \label{eq:Grho}\\
    C_\rho^0(q,\omega) &= 2\lambda_0^{-1}w_0 f_0\,q^2\,\bigl|G_\rho^0(q,\omega)\bigr|^2\,.
    \label{eq:Crho0}
\end{align}
The factor $q^2$ in $C_\rho^0$ reflects the conserved-density origin of the noise: in the original Langevin equation~\eqref{eq:deltarho} the stochastic current is the divergence of a white noise, so the density noise power spectrum is $\propto q^2$ in Fourier space. In the small $w_0$ limit, the density propagator reduces to the fast-diffusive form $G_\rho^0 = q^{-2}$ and $C_\rho^0 = 2\lambda_{0}^{-1}f_{0}w_{0}q^{-2}$ at low frequency — reproducing in Fourier space the effective noise correlator~\eqref{eq:effnoise} obtained by adiabatic elimination.

Perturbation theory is then constructed by expanding in $\mathcal{L}_I'$ the path integral weight $\eu^{-S}=\eu^{-S_{0}}\sum (-S_I')^{n}/n!$. In this way, the path integrals are Gaussian, and can be computed exactly. Each path integral will be represented as a Feynman diagram, with the rules discussed in App.~\ref{app:feynmanrules}.

\subsection{Tree-level vertex functions}\label{sec:Gamma0}
The natural objects for the perturbative expansion and the renormalization program are the one-particle irreducible (1PI) vertex functions $\Gamma^{n}$ \cite{binney1992theory}. Their generating function $\Gamma[\Phi]$ is the effective potential of the theory, which expresses the statistical weight of a given configuration with $\Phi_{\phi}=\langle\phi\rangle$ as $e^{-\Gamma[\Phi]}$. The $n$-point vertex functions are given by
\begin{equation}
    \begin{split}
        \frac{\delta^n \Gamma[\Phi]}{\delta\Phi_{\phi_1}(q_1)\cdots\delta\Phi_{\phi_n}(q_n)}
        \Bigg|_{\Phi=0}=\\
        =\delta\left(\sum_{i} q_{i}\right)\Gamma_{\phi_1\cdots\phi_n}&(q_1,\ldots,q_{n-1})
    \end{split}
    \label{eq:Gamma_def}
\end{equation}

Diagrammatically, $\Gamma$ vertex functions receive contributions exclusively from \emph{amputated 1PI diagrams}: diagrams that remain connected after the removal of any single internal line, with the external-leg propagators stripped off. At tree level, the nonvanishing 1PI functions read
\begin{align}
    \Gamma_{\psi\hpsi}^0(q,\omega) &= \bigl[G_\psi^0\bigr]^{-1} =
        -\iu\lambda_0^{-1}\omega + q^2 + \tau_0\,,
    \label{eq:Gamma_pp}\\
    \Gamma_{\hpsi\hpsi}^0(q,\omega) &= \lambda_0^{-1}\,,
    \label{eq:Gamma_phph}\\
    \Gamma_{\rho\hrho}^0(q,\omega) &= \bigl[G_\rho^0\bigr]^{-1} =
        -\iu\lambda_0^{-1}w_0\,\omega + q^2\,,
    \label{eq:Gamma_rr}\\
    \Gamma_{\hrho\hrho}^0(q,\omega) &= \lambda_0^{-1}w_0 f_0\,q^2\,,
    \label{eq:Gamma_rhrh}
\end{align}
along with the two interaction vertices
\begin{equation}
    \Gamma_{\psi\psi\psi\hpsi}^0 = -u_0\,, \qquad
    \Gamma_{\rho\psi\hpsi}^0 = -1\,.
    \label{eq:Gamma_vertices}
\end{equation}
The two-point 1PI functions $\Gamma_{\psi\hpsi}$ and $\Gamma_{\rho\hrho}$ are simply the inverse propagators: the relations $\Gamma_{\psi\hpsi}\,G_\psi = 1$ and $\Gamma_{\rho\hrho}\,G_\rho = 1$ hold at all orders in perturbation theory and are the basis of Dyson's equation. The mixed noise vertices $\Gamma_{\hpsi\hpsi}^0$ and $\Gamma_{\hrho\hrho}^0$ are not propagators but rather encode the strength of the stochastic forcing.  The four-point vertex $\Gamma_{\psi\psi\psi\hpsi}^0 = -u_0$ and the three-point vertex $\Gamma_{\rho\psi\hpsi}^0 = -1$ are the diagrammatic vertices, corresponding to the non-linearities in the equations of motion. The rescaled form $\Gamma_{\rho\psi\hpsi}^0 = -1$, rather than $-g_0$, reflects the field redefinition of $\delta\rho$ in Eq.~\eqref{eq:rescale_rho}, which absorbed $g_0$ into the field normalization.

\subsection{Power counting and the upper critical dimension}\label{sec:powercounting}
To identify which 1PI functions require renormalization it is sufficient to perform a power-counting analysis. Vertex functions with a non-negative canonical dimension might contain superficially divergent diagrams in their expansion, whose removal requires the introduction of counterterms. Vertex functions with negative canonical dimension, on the other hand, are superficially convergent. Divergences might still arise in their perturbative expansion, but they are removed once the superficially divergent vertex functions are properly renormalized. No counterterms for these vertex functions are thus required. The degree of divergence of a vertex function is determined by its canonical dimension, namely
\begin{equation}
    d_{\Gamma} = d + z - n_{\phi}d_{\phi}\,,
    \label{eq:canonical_dim}
\end{equation}
where $n_{\phi}$ is the number of fields $\phi$ in the vertex function and $d_{\phi}$ is the canonical dimension of the field $\phi$. 

The canonical dimensions of the fields are determined by requiring that each term in the (rescaled) action $\mathcal{L}_0' + \mathcal{L}_I'$ is dimensionless. Working in units of momentum ($d_{q}=1$) and frequency ($d_{\omega}=z$), the canonical dimensions of the fields are
\begin{align}
    d_{\psi} &= \frac{d-2}{2}\,, &
    d_{\hpsi} &= \frac{d+2z-2}{2}\, \\
    d_{\rho} &= 2\,,&
    d_{\hrho} &= d+z-4\,,
    \label{eq:field_dims}
\end{align}
and those of the coupling constants are
\begin{align}
    d_{\lambda_0} &= z-2&
    d_{w_0} &= 0&
    d_{u_0} &= d_{f_0} = \epsilon&
    d_{r_0} &= 2
    \label{eq:coupling_dims}
\end{align}
Both $u_0$ and $f_0$ carry canonical dimension $\epsilon\equiv 4-d$: they are marginal at $d_c=4$, relevant below, and irrelevant above. The timescale ratio $w_0$ is dimensionless at all $d$, consistent with its role as a crossover parameter rather than a relevant coupling. The upper critical dimension, at which the theory is logarithmically divergent, is therefore $d_c=4$. Above $d_{c}$ the theory is UV-divergent, while below $d_{c}$ it is IR-divergent. The IR-divergences are associated with the emergence of anomalous scaling of the correlation functions at the critical point. Above $d_{c}$, the critical exponents take their mean-field values, and the Gaussian theory becomes asymptotically exact at large scales. The critical exponents characterizing the scaling can thus be computed in a controlled way by performing an expansion in $\epsilon$ around the upper critical dimension, a standard tool of the theory of critical phenomena~\cite{peskin1995introduction,amit2005fieldtheory,binney1992theory}. The renormalization program consists in systematically removing the UV divergences at each order in perturbation theory, and then using the resulting renormalized theory to compute the critical exponents as a power series in $\epsilon$.

\section{Perturbative renormalization}
\label{sec:RG1}
The superficially divergent 1PI functions — those whose canonical dimension is non-negative and require counterterms --- are $\Gamma_{\psi\hpsi}$, $\Gamma_{\hpsi\hpsi}$, $\Gamma_{\rho\hrho}$, $\Gamma_{\hrho\hrho}$, $\Gamma_{\psi\psi\psi\hpsi}$, and $\Gamma_{\rho\psi\hpsi}$. The vertex function $\Gamma_{\hrho\psi\psi}$ also has a non-negative canonical dimension on paper, but it vanishes at all orders due to the absence of an interaction mixing $\hrho$ and $\psi$. This property follows from the structural absence of feedback from the order parameter $\psi$ to the density $\rho$, a defining property of the model, and hence $\Gamma_{\hrho\psi\psi}=0$ is protected by this effective symmetry. Ensuring the finiteness of these functions at each order in perturbation theory is the basis of the renormalization program. In practice, divergences arise in the following nine quantities
\begin{equation}
\begin{split}
    &\Gamma_{\psi\hpsi}(0,0)\,,\quad
     \partial_\omega\Gamma_{\psi\hpsi}(0,\omega)\big|_{\omega=0}\,,\quad
     \partial_{q^2}\Gamma_{\psi\hpsi}(q,0)\big|_{q=0}\,,\\
    &\Gamma_{\hpsi\hpsi}(0,0)\,,\quad
     \partial_\omega\Gamma_{\rho\hrho}(0,\omega)\big|_{\omega=0}\,,\quad
     \partial_{q^2}\Gamma_{\rho\hrho}(q,0)\big|_{q=0}\,,\\
    &\partial_{q^2}\Gamma_{\hrho\hrho}(q,0)\big|_{q=0}\,,\quad
     \Gamma_{\psi\psi\psi\hpsi}(\bol{0})\,,\quad
     \Gamma_{\rho\psi\hpsi}(\bol{0})\,.
\end{split}
\label{eq:divs}
\end{equation}
All higher order derivatives of the 1PI functions with respect to frequency and momentum are superficially convergent and do not require renormalization. Divergences in the three quantities in the first line can be reabsorbed into a renormalization of the mass, kinetic coefficient, and amplitude of the $\psi$ field, respectively; the next three require renormalization of the noise amplitude of $\psi$, the diffusion coefficient and amplitude of the $\rho$ field; while the last three lead to the renormalization of the noise amplitude of the density field $\rho$ and the two interaction vertices.


\subsection{Renormalization factors}
\label{sec:gamma}
To take care of all nine divergences in the $\Gamma$ functions listed in Eq.~\eqref{eq:divs}, we introduce a multiplicative renormalization of the fields and parameters of the theory~\cite{tauber2014criticaldynamics}, in addition to the additive renormalization of the mass $r$. The renormalized fields $\psi_R$, $\hpsi_R$, $\rho_R$, and $\hrho_R$ are related to the bare fields by
\begin{align}
    \psi_R &= Z_\psi^{1/2}\,\psi_0\,, &
    \hpsi_R &= Z_{\hpsi}^{1/2}\,\hpsi_0\,,\\
    \rho_R &= Z_\rho^{1/2}\,\rho_0\,,&
    \hrho_R &= Z_{\hrho}^{1/2}\,\hrho_0\,.
\end{align}
The renormalized couplings are defined by
\begin{align}
	u_R &= Z_u u_0K_{d}\mu^{-\epsilon} &
	f_R &= Z_f f_0K_{d}\mu^{-\epsilon} \\
    w_R &= Z_w w_0
	\label{eq:renormalized_couplings}
\end{align}
where $\mu$ is the renormalization scale, $K_d\equiv S_d/(2\pi)^d$ with $S_d$ being the surface area of the unit sphere in $d$ dimensions, and the $Z$-factors absorb UV divergences order by order in perturbation theory. The mass requires both an additive and a multiplicative renormalization. In particular, as previously anticipated, fluctuations shift the critical point from $r_{0}=0$ to $r_{0}=r_{c}$. Additive renormalization of the mass is thus required to reabsorb this shift, and we write $r_0 = r_c + \tau_0$, with $\tau_0$ being the distance from criticality. The multiplicative renormalization of the mass is then performed by defining
\begin{equation}
    \tau_R = Z_\tau \tau_0\mu^{-2}
\end{equation}
Finally, the kinetic coefficient $\lambda$ also requires a multiplicative renormalization, which we write as
\begin{equation}
    \lambda_R = Z_\lambda \lambda_0\,.
\end{equation}

We work in the minimal subtraction scheme throughout, requiring the renormalization constants to absorb only the divergent contributions to $\Gamma$. The UV poles arising in each Gamma function are cancelled at each order in perturbation theory. The renormalization point $\mu$ is chosen in such a way to avoid IR divergences. Here, we ensure this by requiring $\tau_{R}=1$, which at lowest order corresponds to $\tau\simeq\mu^{2}$.

Renormalization of the $\psi$ sector ($\Gamma_{\psi\hpsi}$, $\Gamma_{\hpsi\hpsi}$ and $\Gamma_{\psi\psi\psi\hpsi}$) fixes five independent $Z$-factors ($Z_\psi,\,Z_{\hpsi},\,Z_{\lambda},\,Z_{\tau},\,Z_{u}$). The factor $Z_{\psi}$ can then be fixed by the renormalization of the $\Gamma_{\rho\psi\hpsi}$ vertex, which is the only vertex mixing $\psi$ and $\rho$. Finally, the remaining $Z$-factors ($Z_{\hrho},\,Z_{f},\,Z_{w}$) are fixed by the renormalization of the density sector ($\Gamma_{\rho\hrho}$, and $\Gamma_{\hrho\hrho}$). The explicit derivation of the $Z$ factors — including the identification of the contributing diagrams and the evaluation of the loop integrals — is given in Appendix~\ref{app:oneloop}. It is worth noting that, to one-loop order, the density sector is not renormalized, leading to the relations $Z_{\hrho}=Z_{\rho}^{-1}$, $Z_w = Z_\lambda$, and $Z_f = Z_\rho$. As we shall discuss later in Sec.~\ref{sec:exact}, this result is not an artifact of perturbation theory but follows from the absence of RG-relevant interactions in the density sector, and thus holds to all orders in perturbation theory.

\subsection{The RG equations}
The renormalized vertex functions are thus given by~\cite{tauber2014criticaldynamics}
\begin{equation}
\begin{split}
    &\Gamma_{\psi^n\rho^l\hpsi^k\hrho^p}^R(\{q,\omega\};\{\lambda_{R},\mathrm{p}_R\})=\\
    &\quad=Z_{\psi}^{-\frac n2} Z_{\rho}^{-\frac  l2}Z_{\hpsi}^{-\frac k2} Z_{\hrho}^{-\frac p2} \Gamma_{\psi^n\rho^l\hpsi^k\hrho^p}^{0}(\{q,\omega\};\{\mathrm{p}_0\})
\end{split}
\label{eq:renGamma}
\end{equation}
Here $\mathrm{p}$ are the parameters of the system, namely $\lambda,\tau,w,f,u$. We now aim to understand how these renormalized quantities depend on the observation scale, namely the normalization point $\mu$. The bare vertex functions $\Gamma_{\psi^n\rho^l\hpsi^k\hrho^p}^{0}(\{q,\omega\};\{\mathrm{p}_0\})$ do not depend on the normalization scale $\mu$, hence
\begin{equation}
    \mu\frac{d}{d\mu}\Gamma_{\psi^n\rho^l\hpsi^k\hrho^p}^{0}(\{q,\omega\};\{\mathrm{p}_0\})=0
\end{equation}
Starting from this, and using \eqref{eq:renGamma}, the following RG equation for the renormalized $\Gamma^R$ can be derived
\begin{equation}
    \left[\mu\frac{\partial}{\partial\mu}+\frac{n_{\phi}}{2}\gamma_{\phi}-\beta_{\mathrm{p}}\partial_{\mathrm{p}}+\gamma_{\lambda}\partial_{\lambda}\right]\Gamma^R=0
    \label{eq:theRGeq}
\end{equation}
Summation over all fields $\phi=\{\psi,\rho,\hpsi,\hrho\}$ and all effective parameters $\mathrm{p}$ is understood.
Here the $\gamma$ function for a given quantity $a$ is
\begin{equation}
    \gamma_{a}=\left.\mu \frac{d}{d\mu}\right|_0 \log{Z_{a}}
\end{equation}
with the subscript $0$ meaning that derivatives are taken at fixed bare parameters. Effective parameters $\tau,w,f,u$ are instead characterized by their $\beta$ functions, given by
\begin{equation}
    \beta_{\mathrm{p}} =-\mu\left.\frac{d}{d\mu}\right|_0 \mathrm{p}_R=\mathrm{p}_R\left(d_{\mathrm{p}}-\gamma_{\mathrm{p}}\right)
\end{equation}
with $d_{\mathrm{p}}$ being the canonical dimension of $\mathrm{p}$ and $\gamma_{\mathrm{p}}$ the anomalous dimension of $\mathrm{p}$.

\subsection{One-loop renormalization}
At the critical point $\tau_{R}^{*}=0$, explicit evaluation of the $\gamma$ functions at one-loop level leads to
\begin{align}
    \gamma_{\psi}&=\frac{w_{R} }{(1+w_{R})^3} f_{R}\, , \\
    \gamma_{\hpsi}&=- \frac{w_{R}(1+2w_{R})}{(1+w_{R})^3} f_{R}\, , \\
    \gamma_{\rho}&=-2\frac{w_{R}^3 }{(1+w_{R})^3} f_{R} + 6 u_{R}\,,\\
    \gamma_{u}&=9 u_{R} - w_{R} \frac{4+8w_R+6w_R^2}{(1+w_R)^3} f_{R}\, , \\
    \gamma_{\lambda}&= \frac{w_{R}^3 }{(1+w_{R})^3} f_{R}\,, \\
    \gamma_{\tau}&=3u_{R} - \frac{w_{R}^3 }{(1+w_{R})^3} f_{R}\,.
\end{align}
Other $\gamma$ functions follow from the relation between $Z$ factors, which gives $\gamma_{\hrho}=-\gamma_{\rho}$, $\gamma_{w}=\gamma_{\lambda}$, and $\gamma_{f}=\gamma_{\rho}$. The $\beta$ functions for the effective parameters follow from $\beta_{\mathrm{p}}=(d_{\mathrm{p}}-\gamma_{\mathrm{p}})$ and are given by:
\begin{align}
    \beta_u &= u_R\!\left(\epsilon - 9u_R
        + w_R\frac{4+8w_R+6w_R^2}{(1+w_R)^3}\,f_R\right),
        \label{eq:betau_wf}\\
    \beta_f &= f_R\!\left(\epsilon - 6u_R
        + \frac{2w_R^3}{(1+w_R)^3}\,f_R\right),
        \label{eq:betaf_wf}\\
    \beta_w &= -\frac{w_R^4}{(1+w_R)^3}\,f_R\,.
        \label{eq:betaw_wf}
\end{align}
Also $\tau$, being a dimensionful effective parameter, has its own $\beta$ function. Since we are working in the vicinity of the critical point $\tau_{R}=0$, we evaluate its $\beta$ function only to first order in $\tau$, which leads to
\begin{equation}
    \beta_{\tau}=\tau_{R}\left(2-3u_{R} + \frac{w_{R}^3 }{(1+w_{R})^3} f_{R}\right)\,.
\end{equation}

\subsection{RG flow equations and the BIM parameterization}
\label{sec:betaBIM}
The RG equation \eqref{eq:theRGeq} can be solved using the method of characteristics \cite{peskin1995introduction}: a change in renormalization scale $\mu\to e^{-l}\mu$ can be accounted for by a change of the effective parameters, which evolve according to
\begin{equation}
    \dot{\mathrm{p}}_R = \beta_{\mathrm{p}}(\mathrm{p}_R)\,,
\end{equation}
and a change of the amplitudes. This defines a flow in the parameter space, which is the basis of the RG analysis. The critical behavior is controlled by the fixed points of this flow, which are defined by $\beta_{\mathrm{p}}=0$ for all effective parameters $\mathrm{p}$. The critical exponents are then extracted from the linearization of the flow around the fixed point.

The flow defined by Eqs.~\eqref{eq:betau_wf}--\eqref{eq:betaw_wf} appears to have no infrared-stable fixed point: the only nontrivial solution with $f_R^*=0$ is the Model A point $u_R^*=\epsilon/9$, which is unstable to perturbations in $f_R$. Once $f_R\neq0$, the coupling $f_R$ grows without bound under the RG flow. However, the system is simultaneously driven toward $w_R\to0$, so that the correct infrared coupling is not $f_R$ itself but the combination $w_R f_R$, which remains finite along the flow. This precise combination is not a coincidence: it is the prefactor of the density correlations in the fast-diffusion limit, and thus controls the strength of the coupling between the order parameter and the density fluctuations.

\begin{table}[t]
\centering
\begin{tabular}{r|c|c|c|c}
\hline\hline
FP & $u_R^*$ & $\tif_R^*$ & $X_R^*$ & Unstable \\ [0.5ex]
\hline
G        & $0$                        & $0$                       & ---              & $2$ \\[1ex]
Model A  & $\dfrac{\epsilon}{9}$      & $0$                       & $0$              & $1$ \\[1ex]
Quenched\,\cite{grinstein1977dynamics}
	 & $\sqrt{{8/159}\epsilon}$ & $\sqrt{{6/53\epsilon}}$ & $1$     & $1$ \\[1ex]
BIM      & $\dfrac{\epsilon}{6}$      & $\dfrac{\epsilon}{8}$     & $0$              & $0$ \\[1ex]
\hline\hline
\end{tabular}
\caption{Fixed points of the one-loop RG flow in the coupling-constant space $(u_R,\tif_R,X_R)$. The last column gives the number of unstable (relevant) directions on the critical manifold $\tau_R^*=0$.
The Gaussian fixed point controls the mean-field regime $d>4$; below $d_c=4$ it is unstable in both directions.
The Model A fixed point reproduces Ising criticality ($\tif_R^*=0$) but is unstable to density-coupling perturbations.
The Quenched fixed point ($X_R^*=1$) governs the diluted-Ising universality class; its coordinates are determined at two loops and taken from Ref.~\cite{grinstein1977dynamics}.
The BIM fixed point is IR-stable at one loop.}
\label{tab:FP}
\end{table}

This observation motivates introducing a new set of coupling constants
\begin{align}
    X_R &\equiv \frac{w_R}{1+w_R}\,, &
    \tif_R &\equiv X_R f_R\,,
    \label{eq:BIM_vars}
\end{align}
The variable $X_R$ remains finite in both the fast-diffusion limit ($w_R\to0$, $X_R\to0$) and the quenched limit ($w_R\to\infty$, $X_R\to1$). The coupling $\tif_R$ similarly interpolates between the two regimes: $\tif_R\approx w_R f_R$ for small $w_R$, while $\tif_R\approx f_R$ for large $w_R$. In terms of these variables the one-loop $\beta$ and $\gamma$-functions read
\begin{align}
    \beta_{\tif} &= \tif_R\!\left(\epsilon - 6u_R
        + \tif_R\,X_R^2(1+X_R)\right),
        \label{eq:betaftil}\\
    \beta_u &= u_R\!\left(\epsilon - 9u_R
        + 2\tif_R(X_R^2+2)\right),
        \label{eq:betau}\\
    \beta_X &= -\tif_R\,X_R^3(1-X_R)\,.
        \label{eq:betaX}
\end{align}
\begin{align}
    \gamma_{\psi}    &= \tif_R(1-X_R)^2\,,
        \label{eq:gamma_psi_BIM}\\
    \gamma_{\hpsi}   &= -\tif_R(1-X_R^2)\,,
        \label{eq:gamma_hpsi_BIM}\\
    \gamma_{\rho}   &= -2 X_R^2\,\tif_R+6u_{R}\,,
        \label{eq:gamma_rho_BIM}\\
    \gamma_{\lambda}  &= X_R^2\,\tif_R\,,
        \label{eq:gamma_lambda_BIM}\\
    \gamma_{\tau}    &= 3u_R - X_R^2\,\tif_R\,.
        \label{eq:gamma_tau_BIM}
\end{align}

\subsection{Fixed points and RG flow}
\label{sec:fixedpoints}
The fixed points of the RG flow are determined by setting $\beta_{\tif}=\beta_u=\beta_X=0$ on the critical manifold $\tau_R^*=0$. The solutions are collected in Table~\ref{tab:FP}.

The \emph{Gaussian} fixed point ($u_R^*=\tif_R^*=0$) controls the critical behavior for $d>4$; below $d_c=4$ it is unstable in both the $u_R$ and $\tif_R$ directions. The value of $X_R^*$ is undefined at this point, since $\beta_X$ vanishes trivially when $\tif_R=0$.

The \emph{Model A} fixed point ($\tif_R^*=0$, $u_R^*=\epsilon/9$) reproduces the Ising universality class with purely relaxational dynamics. The condition $\beta_X=0$ is satisfied for any value of $X_R^*$ at one loop. The two-loop corrections select $X_R^*=0$ as the dynamically stable value, as a consequence of $z>2$ (see Sec.~\ref{sec:exact}). This fixed point carries one unstable direction, corresponding to perturbations in $\tif_R$.

The \emph{Quenched} fixed point ($X_R^*=1$) is reached in the limit of immobile disorder ($X_R\to1$) and is analyzed separately in Sec.~\ref{sec:quenched}. Its coordinates in the $(u_R,\tif_R)$ plane require a two-loop calculation and are taken from Ref.~\cite{grinstein1977dynamics}.

\begin{figure}[t]
\includegraphics[width=\columnwidth]{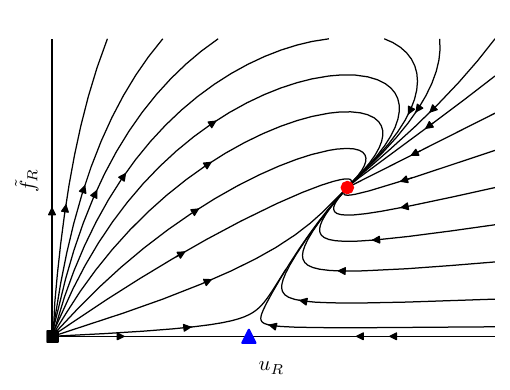}
\caption{One-loop RG flow in the $(u_R,\tif_R)$ coupling plane at $X_R=0$. The Gaussian fixed point (black square) is unstable below $d_c=4$. The Model A fixed point (blue triangle), corresponding to Ising criticality, is unstable to the density coupling $\tif_R$. The BIM fixed point (red circle) is the IR-stable fixed point controlling the critical dynamics of the BIM. The flow on this plane is exact at one loop: since $\beta_X\propto X_R^3$, the manifold $X_R=0$ is invariant under the one-loop RG.}
\label{fig:RGflow}
\end{figure}

The novel \emph{BIM} fixed point,
\begin{align}
    X_R^* = 0\,, \qquad u_R^* = \frac{\epsilon}{6}\,, \qquad \tif_R^* = \frac{\epsilon}{8}\,,
    \label{eq:BIM_FP}
\end{align}
is the IR-stable fixed point of the flow and controls the critical dynamics of the BIM near $d=4$. The one-loop RG flow on the $X_R=0$ plane is shown in Fig.~\ref{fig:RGflow}.

\subsection{Critical exponents and stability}
\label{sec:exponents}
The four critical exponents are related to the anomalous dimensions evaluated at the fixed point of interest. Direct inspection of the RG equations gives
\begin{align}
    \nu &= \frac{1}{2-\gamma_\tau^*}\,, &
    z   &= 2 + \gamma_\lambda^*\,, \nonumber\\
    \eta  &= -\gamma_\psi^*\,, &
    \eta' &= -\frac{\gamma_\psi^* + \gamma_{\hpsi}^*}{2}\,.
    \label{eq:exp_def}
\end{align}
Here, $\eta'=2-\gamma/\nu$, with $\gamma$ being the standard susceptibility exponent, related to the response to external perturbations.

Substituting the BIM fixed-point values $X_R^*=0$, $u_R^*=\epsilon/6$, $\tif_R^*=\epsilon/8$ into Eqs.~\eqref{eq:gamma_psi_BIM}--\eqref{eq:gamma_tau_BIM} yields
\begin{align}
    \nu   &= \frac{1}{2} + \frac{\epsilon}{8} + \mathcal{O}(\epsilon^2)\,, &
    z     &= 2 + \mathcal{O}(\epsilon^2)\,, \nonumber\\
    \eta  &= -\frac{\epsilon}{8} + \mathcal{O}(\epsilon^2)\,, &
    \eta' &= \mathcal{O}(\epsilon^2)\,.
    \label{eq:BIM_exp}
\end{align}
Several features of these results distinguish the BIM universality class from the Ising one already at one loop. The correlation-length exponent differs from the Ising value $\nu_\mathrm{Ising}=\frac{1}{2}+\frac{\epsilon}{12}+\mathcal{O}(\epsilon^2)$.
More significantly, $\eta\neq\eta'$: in equilibrium dynamics, a Ward identity follows from the fluctuation-dissipation theorem, and forces $\gamma_{\psi}=\gamma_{\hpsi}$, so that these two exponents are equal~\cite{janssen1976on,tauber2014criticaldynamics}. Their splitting is thus a direct signature of the non-equilibrium nature of the BIM phase transition. Note also that $\eta$ is of order $\epsilon$, and not $\epsilon^2$ as happens in Model A.

The negative sign of $\eta$ is also quite peculiar, and reflects the fact that the BIM fixed point lies at $X_R^*=0$, i.e., in the fast-diffusion limit. As discussed in Sec.~\ref{sec:fast_diff}, in this limit the coupling to diffusion generates an effective noise with long-range spatial correlations, which enhances order-parameter fluctuations and drives $\eta<0$.

\subsubsection*{Stability of the BIM fixed point}

Although the BIM fixed point is IR-stable with respect to the two-coupling flow in the $X_R=0$ plane, the full three-dimensional stability must be examined. We compute the eigenvalues and eigenvectors of the stability matrix $M_{ij}=\partial\beta_i/\partial g_j\big|_{g^*}$ on the critical manifold, with $(g_1,g_2,g_3)=(X_R,u_R,\tif_R)$. At leading order in $\epsilon$,
\begin{align}
    e_1 &= 0\,, & \mathbf{v}_1 &= (1,\,0,\,0)\,, \label{eq:e1}\\
    e_2 &= -\frac{\epsilon}{2}\,, & \mathbf{v}_2 &= (0,\,2,\,3)\,, \label{eq:e2}\\
    e_3 &= -\epsilon\,, & \mathbf{v}_3 &= (0,\,4,\,3)\,. \label{eq:e3}
\end{align}
The eigenvalues $e_2$ and $e_3$ are both negative, confirming stability in the $(u_R,\tif_R)$ subspace. The marginal direction $e_1=0$ corresponds to perturbations along $X_R$, and reflects the fact that $\beta_X\propto X_R^3$: convergence toward $X_R^*=0$ is slower than exponential.

The correction-to-scaling exponent $\omega$, which governs how rapidly observables approach their asymptotic power laws, is the magnitude of the largest (least negative) eigenvalue. Since $e_1=0$ at one loop,
\begin{equation}
    \omega = \mathcal{O}(\epsilon^2)\,.
    \label{eq:omega}
\end{equation}
For comparison, $\omega=\epsilon$ at the Ising fixed point at one loop.
A vanishing $\omega$ at one loop raises the question of whether higher-order corrections could destabilize the BIM fixed point. In Sec.~\ref{sec:two_loop} we compute the two-loop contribution to $e_1$ and show that it is negative, confirming stability.

\subsection{Two-loop renormalization}\label{sec:two_loop}
At one loop, for the BIM fixed point we found $\gamma_\lambda^*=0$ and $\gamma_\psi^*+\gamma_{\hpsi}^*=0$, so that $z=2$ and $\eta'=0$ to this order. These vanishings left the stability of the fixed point in question through the marginal eigenvalue $e_1=0$. Here we compute the leading corrections to $z$ and $\eta'$, which arise at two loops, and use them to settle the stability question.

A full two-loop calculation of all $Z$-factors would require evaluating a large number of diagrams. However, as in the analogous treatment of Model A~\cite{hohenberg1977theory}, the fact that $z=2$ and $\eta'=0$ at one loop means that the leading $\epsilon^2$ corrections to both exponents are determined solely by $Z_\lambda$ and $Z_{\hpsi}Z_\psi$, without requiring the full two-loop renormalization of every coupling. Specifically, it suffices to use the one-loop fixed-point values of $u_R$ and $\tif_R$ as inputs and compute only the two-loop contributions to these two $Z$-factor combinations at $X_R=0$.

\begin{figure}
    \begin{overpic}[width=0.5\textwidth, page=6]{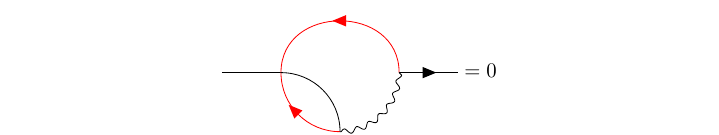}
        \put(0, 18){(A)}
        \put(35, 18){(B)}
        \put(70, 18){(C)}
    \end{overpic}
\caption{\textbf{Two-loop diagrams} contributing to the vertex function $\Gamma_{\psi \hpsi}$. At $w_{0}=0$, the correlator $C_{\rho}$ is delta-correlated in time, and all the diagrams with closed loops of propagators, such as (A), are exactly zero because of causality --- in red the closed oriented loop that cancels the diagram for causality. The only two non-zero 1PI diagrams are the sunset diagram (B), also present in Model A \cite{tauber2014criticaldynamics, hohenberg1977theory}, which gives a contribution of order $u_{R}^2$ and diagram (C) which gives a contribution of order $\tif_{R} u_{R}$. }
\label{fig:twoloop}
\end{figure}

A further simplification follows from causality. The causal structure of the MSR action forbids closed loops of retarded propagators, which eliminates most two-loop diagrams contributing to $\Gamma_{\psi\hpsi}$ when evaluated at $X_R^*=0$. Fig~\ref{fig:twoloop}(A) shows an example of such diagrams; only two diagrams survive, shown in Fig~\ref{fig:twoloop}(B-C). Their evaluation yields the following two-loop anomalous dimensions at $X_R=0$:
\begin{align}
    \gamma_\lambda &= \frac{u_R}{2}\!\left(
        -\tif_R - 3u_R + 18\log\!\tfrac{4}{3}\, u_R
        \right),
    \label{eq:gamma_lambda_2loop}\\
    \gamma_\psi + \gamma_{\hpsi} &= -u_R\!\left(
        \tif_R\!\left(1 - 12\log\!\tfrac{4}{3}\right) + 3u_R
        \right).
    \label{eq:gamma_sum_2loop}
\end{align}
As a consistency check, substituting the Model A fixed-point values $u_R^*=\epsilon/9$, $\tif_R^*=0$ reproduces the standard results $z = 2+\frac{\epsilon^2}{54}(6\log\frac{4}{3}-1)$ and $\eta=\eta'=\epsilon^2/54$~\cite{hohenberg1977theory}.

At the BIM fixed point $u_R^*=\epsilon/6$, $\tif_R^*=\epsilon/8$, Eqs.~\eqref{eq:gamma_lambda_2loop} and~\eqref{eq:gamma_sum_2loop} give
\begin{align}
    z &= 2 + \frac{\epsilon^2}{96}\!\left(24\log\!\tfrac{4}{3} - 5\right),
    \label{eq:z_2loop}\\
    \eta' &= \frac{\epsilon^2}{96}\!\left(5 - 12\log\!\tfrac{4}{3}\right).
    \label{eq:etap_2loop}
\end{align}
Since $24\log(4/3)-5 \approx 1.90 > 0$, we have $\gamma_\lambda^*>0$ and hence $z>2$ at the BIM fixed point. Note also that, while $\eta<0$ at one-loop, the response-function exponent $\eta'>0$ at two loops. A summary of the critical exponents at the three perturbatively accessible fixed points, to lowest nontrivial order in $\epsilon$, is given in Table~\ref{tab:exp}.

\begin{table*}[t]
\centering
\setlength\extrarowheight{4pt}
\begin{tabular*}{\textwidth}{@{\extracolsep{\fill}}r|ccccc}
\hline\hline
FP & $\nu$ & $\eta$ & $\eta'$ & $z$ & $\omega$ \\[4pt]
\hline
G       & $\dfrac{1}{2}$ & $0$ & $0$ & $2$ & $0$ \\[4pt]
Model A & $\dfrac{1}{2}+\dfrac{\epsilon}{12}$ & $\dfrac{\epsilon^2}{54}$ &
          $\dfrac{\epsilon^2}{54}$ &
          $2+\dfrac{\epsilon^2}{54}\!\left(6\log\tfrac{4}{3}-1\right)$ &
          $\epsilon$  \\[4pt]
BIM     & $\dfrac{1}{2}+\dfrac{\epsilon}{8}$ & $-\dfrac{\epsilon}{8}$ &
          $\dfrac{\epsilon^2}{96}\!\left(5-12\log\tfrac{4}{3}\right)$ &
          $2+\dfrac{\epsilon^2}{96}\!\left(24\log\tfrac{4}{3}-5\right)$ &
          $\dfrac{\epsilon^2}{96}\!\left(24\log\tfrac{4}{3}-5\right)$ \\[4pt]
\hline\hline
\end{tabular*}
\caption{Critical exponents at the Gaussian (G), Model A, and BIM fixed points, to lowest nontrivial order in $\epsilon=4-d$. The exponent $\omega$ is the correction-to-scaling exponent, equal in magnitude to the least-negative eigenvalue of the stability matrix restricted to the critical manifold. All exponents are given at the lowest nontrivial order in $\epsilon$. At the BIM fixed point, $\nu$ and $\eta$ are the one-loop exponents; while anomalous corrections to $\eta'$, $z$, and $\omega$ first appear at two loops.}
\label{tab:exp}
\end{table*}

An exact relation $Z_w=Z_\lambda$ (see Sec.~\ref{sec:exact}) implies $\beta_w = -w_R\gamma_\lambda$, and hence
\begin{equation}
    \beta_X = X_R(X_R-1)\,\gamma_\lambda\,.
    \label{eq:betaX_exact}
\end{equation}
At $X_R^{*}=0$, the stability-matrix thus reads
\begin{align}
    \partial_{u_R}\beta_X\big|_{X_R=0} &= 0\,, &
    \partial_{\tif_R}\beta_X\big|_{X_R=0} &= 0\,, \\
     \partial_{X_R}\beta_X\big|_{X_R=0}& = -\gamma_\lambda^*\,.
\end{align}
Hence, $-\gamma_\lambda^*$ will always be one of the eigenvalues of the stability matrix. At one loop $\gamma_\lambda^*=0$, yielding the marginal eigenvalue $e_1=0$ found in Sec.~\ref{sec:exponents}. At two loops, however, $\gamma_\lambda^*>0$, so
\begin{equation}
    e_1 = -\gamma_\lambda^* = -\frac{\epsilon^2}{96}\!\left(24\log\!\tfrac{4}{3}-5\right) < 0\,.
    \label{eq:e1_2loop}
\end{equation}
The BIM fixed point is therefore stable in the full coupling-constant space in the vicinity of $d=4$. The correction-to-scaling exponent is
\begin{equation}
    \omega = \gamma_\lambda^* = \frac{\epsilon^2}{96}\!\left(24\log\!\tfrac{4}{3}-5\right),
    \label{eq:omega_2loop}
\end{equation}
which evaluates to $\omega\approx 0.020$ at $d=3$ ($\epsilon=1$). This small value indicates that, despite the convergence being exponential, it may be very slow in practice. According to finite-size scaling theory~\cite{cardy1988finite}, all singular observables acquire corrections governed by $\omega$. For instance, the order-parameter susceptibility takes the form
\begin{equation}
    \chi(L, \tau) = L^{2-\eta}\, f_{\chi}\!\left(\tau L^{1/\nu}\right)
        \left(1 + A\, L^{-\omega}\right),
    \label{eq:fss_correction}
\end{equation}
A small $\omega$ implies that very large systems are required to reach the asymptotic scaling regime, and that corrections to asymptotic scaling can be observed only on systems spanning many decades in $L$.

\subsection{The quenched-disorder limit}
\label{sec:quenched}
The BIM contains a well-defined quenched-disorder limit that connects it to the diluted Ising universality class. Here we derive the renormalization-group equations appropriate to this limit, explain why the $\beta$-functions of Sec.~\ref{sec:betaBIM} do not reproduce them, and briefly discuss the instability of the resulting fixed point with respect to finite defect motility.

When $w_0(q/\Lambda)^{z-2}\gg1$, the density field relaxes on timescales much longer than those of the order parameter: $\tau_\rho\gg\tau_\psi$. From the perspective of the order-parameter dynamics, $\rho$ is effectively frozen. In rescaled variables ($\rho_0\to\tif_0$; see Sec.~\ref{sec:rescaling}), the two-point correlator~\eqref{eq:Crho0} approaches $\tif_0\,\delta(\omega)$ in the limit $w_0\to\infty$, which in real space corresponds to
\begin{equation}
    \langle\rho(\mathbf{x},t)\rho(\mathbf{y},t')\rangle=\tif_0\,\delta^{(d)}(\mathbf{x}-\mathbf{y})\,,
\end{equation}
signaling a time-independent (quenched) disorder field.

The quenched field $\rho$ can be integrated out exactly. The resulting effective action density for $\psi$ alone is
\begin{equation}
    \begin{split}
    S_0 &= \int d^{d}x\,dt\,\hpsi(x,t)\left(\lambda_0^{-1}\partial_{t} -\nabla^{2} + r_0\right)\psi(x,t) -\\
    &-\int d^{d}x\,dt\,\lambda_0^{-1}\hpsi(x,t)^{2}\,,
    \end{split}
    \label{eq:S0_quenched}
\end{equation}
\begin{equation}
    \begin{split}
        S_I &= u_0\int d^{d}x\,dt\, \hpsi(x,t) \psi(x,t)^3 -\\
        &- \tfrac{1}{2}\tif_0 \int d^{d}x\,dt dt' \hpsi(x,t) \psi(x,t) \hpsi(x,t') \psi(x,t')
    \end{split}
    \label{eq:SI_quenched}
\end{equation}
The first term in $S_I$ is the usual $\psi^4$ interaction, given by the $u_0$ vertex \eqref{eq:vertex_u}.
The second term in $S_I$, on the other hand, couples two response fields $\hpsi$ through the density propagator $C_\rho$, generating a new vertex $\psi\psi\hpsi\hpsi$ that is absent at finite $w_0$. Because of the quenched field, this vertex couples pairs of $\hpsi$ and $\psi$ at different times, and is therefore nonlocal in time. The presence of this new vertex, and in particular the nonlocal time dependence, modifies the renormalization of the theory in the quenched limit, and leads to a different set of $\beta$-functions than those derived in Sec.~\ref{sec:betaBIM} (see Appendix~\ref{app:quenched}).

\begin{figure}
    \includegraphics[width=0.5\textwidth, page=9]{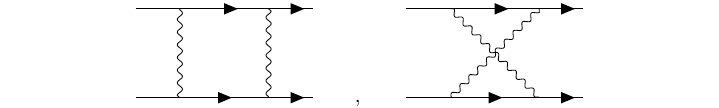}
    \caption{One-loop diagrams contributing to the four-point vertex function $\Gamma_{\psi\psi\hpsi\hpsi}$ in the quenched regime. When the density field is exactly frozen, which happens only when $w_R \to\infty$, two additional one-loop diagrams become UV divergent.}
    \label{fig:quenched_diags}
\end{figure}
At any finite $w_{R}$, $C_{\rho}(\tq)$ is not singular at $\omega_{q}=0$, and the divergences in the vertex $\Gamma_{\psi\psi\hpsi\hpsi}$ are already accounted for by the renormalization of $\Gamma_{\rho\psi\hpsi}$, and therefore do not appear in the list of divergent 1PI functions~\eqref{eq:divs}. In the limit $w_{R}\to\infty$, however, $C_\rho(\mathbf{q},\omega)\to C_\rho(\mathbf{q})\,\delta(\omega)$, and two additional one-loop diagrams contributing to $\Gamma_{\psi\psi\hpsi\hpsi}$ become UV-divergent. These diagrams are shown in Fig.~\ref{fig:quenched_diags}: their contributions must be included in the renormalization of the theory in order to recover the correct behavior in this limit. The fact that different diagrams must be included depending on the value of $w$ is not that unusual in the Callan-Symanzik framework, and it often occurs when considering crossover phenomena. Treating both limits on equal footing would require a simultaneous expansion in $\epsilon$ and $1/w_R$, which is beyond the scope of this work~\cite{FREY199552, cavagna2019renormalization}.

Taking these additional diagrams (see Appendix~\ref{app:quenched}) into account modifies the $\beta$-function for $\tif_R$. The corrected one-loop $\beta$-functions in the quenched limit are
\begin{align}
    \beta_u &= u_R\!\left(\epsilon - 9u_R + 6f_R\right), \label{eq:betau_quenched}\\
    \beta_\tif &= \tif_R\!\left(\epsilon - 6u_R + 4\tif_R\right). \label{eq:betaf_quenched}
\end{align}
These equations are degenerate at one loop: the only solution with $\beta_u=\beta_f=0$ and $u_R,f_R\geq0$ is the Gaussian fixed point, which however is unstable. This degeneracy is a well-known feature of Ising models with quenched impurities~\cite{khmelnitskii1975second,grinstein1976disordered}: the nontrivial fixed point is of order $\sqrt{\epsilon}$ rather than $\epsilon$, and only emerges at two loops. The full two-loop calculation is given in Ref.~\cite{grinstein1976disordered}; the agreement of our one-loop equations with those of the diluted Ising model confirms that the $w_0\to\infty$ limit of the BIM belongs to the diluted-Ising universality class. At the quenched fixed point, $u_R^* = \sqrt{8\epsilon/159}$ and $\tif_R^* = \sqrt{6\epsilon/53}$. The quenched fixed point is characterized by a dynamic exponent~\cite{grinstein1977dynamics}
\begin{equation}
    z = 2 + \sqrt{\frac{6\epsilon}{53}} > 2\,.
    \label{eq:z_quenched}
\end{equation}

As we shall see in Sec.~\ref{sec:exact}, the $\beta$-function for the parameter $X_{R}$ is given by $\beta_{X}=-X_{R}(1-X_{R})\gamma_{\lambda}$, and hence
\begin{equation}
    \partial_{X_{R}}\beta_{X}\big|_{X_{R}=1}=-\gamma_{\lambda}^{*}=-\sqrt{\frac{6\epsilon}{53}}<0\,,
\end{equation}
signaling the instability of the quenched fixed point with respect to finite motility $X_{R}<1$.

\section{Exact results}
\label{sec:exact}

The perturbative results of Sec.~\ref{sec:RG1} establish a novel fixed point close to $d=4$. In this section we derive several relations that hold to all orders in perturbation theory. These allow us to conclude that both the Ising and the diluted-Ising fixed points are unstable for all $d\leq3$, supporting the BIM fixed point as the correct description of the critical behavior also in $d=3$. This conclusion holds, provided no additional fixed point emerges at some intermediate dimension $3<d^*<4$, a remote possibility that perturbation theory however cannot exclude. Furthermore, these results provide an exact relation between the exponents $\nu$ and $z$, which can be used to improve the accuracy of the $\epsilon$-expansion estimates of $\nu$ using the two-loop result for $z$.

\subsection{Relations among renormalization factors}

\subsubsection{Non-renormalization of the density sector}
\label{sec:nonren}
The response field $\hrho$ does not appear in any interaction vertex of the action (Sec.~\ref{sec:propagators}): it couples only bilinearly to $\rho$ through the Gaussian density propagator. Consequently, there are no 1PI diagrams with $\hrho$ as an external leg, and the 1PI functions $\Gamma_{\rho\hrho}$ and $\Gamma_{\hrho\hrho}$ receive no perturbative corrections beyond tree level. The one-loop results $Z_{\hrho}^{1/2}Z_\rho^{1/2}=1$ and $Z_f=Z_\rho$ are therefore exact to all orders:
\begin{align}
    Z_{\hrho} &= Z_\rho^{-1}\,, & Z_f &= Z_\rho\,.
    \label{eq:Zrho_exact}
\end{align}
The same argument applied to $\Gamma_{\hrho\hrho}$ gives $Z_w=Z_\lambda$ exactly,
\begin{equation}
    Z_w = Z_\lambda\,,
    \label{eq:Zw_exact}
\end{equation}
so that the renormalization of $w$ is entirely determined by that of $\lambda$.

\subsubsection{Shift symmetry}
A further exact relation follows from a shift symmetry of the renormalized field theory. The effective action, containing only the relevant operators, is invariant under the simultaneous uniform shift
\begin{align}
    \rho &\;\to\; \rho + \phi\,, & \tau &\;\to\; \tau - \phi\,,
    \label{eq:shift_sym}
\end{align}
for any constant $\phi$. This holds because the coupling between $\psi$ and $\rho$ enters only through the mass term $\tau\hpsi\psi$, and the density field appears in its own equation of motion only through temporal or spatial derivatives. A symmetry of this type also arises in directed percolation coupled to an independently-evolving fluctuating density~\cite{kree1989pollution}.
The symmetry~\eqref{eq:shift_sym} constrains the renormalization factors via $Z_\tau = Z_\rho^{1/2}$. Combined with $Z_f=Z_\rho$, this gives
\begin{equation}
    Z_f = Z_\tau^2\,,
    \label{eq:Zf_exact}
\end{equation}
exact to all orders in perturbation theory. We note that the symmetry~\eqref{eq:shift_sym} is an effective symmetry of the renormalized theory near the BIM fixed point; it is not a symmetry of the microscopic model, and it breaks down at the quenched fixed point ($X_R^*=1$), where the relation $Z_f=Z_\tau^2$ no longer applies.

\subsection{Fixed-point stability}
\subsubsection{Instability of the quenched fixed point}
\label{sec:instab_quenched}

The exact relation~\eqref{eq:Zw_exact} implies that the $\beta$-function for $X_R$ can be written in closed form, see Eq.~\eqref{eq:betaX_exact}. At any fixed point, $z=2+\gamma_\lambda^*$, and $\beta_X^*=0$ requires
\begin{equation}
    \beta_X^* = X_R^*(X_R^*-1)(z-2) = 0\,.
    \label{eq:betaX_FP}
\end{equation}
The three solutions determine the allowed fixed-point values of $z$:
\begin{equation}
    \begin{cases}
        z > 2 & \text{if } X_R^* = 0\,,\\
        z = 2 & \text{if } 0 < X_R^* < 1\,,\\
        z < 2 & \text{if } X_R^* = 1\,.
    \end{cases}
    \label{eq:z_cases}
\end{equation}
The quenched fixed point ($X_R^*=1$) corresponds to the diluted Ising universality class, for which $z>2$ is known both close to $d=4$~\cite{grinstein1977dynamics} and in $d=3$~\cite{hasenbusch2007relaxational}. Since $z>2$ is incompatible with the stability of $X_R^*=1$, Eq.~\eqref{eq:z_cases} shows that the quenched fixed point cannot be a stable fixed point in any dimension $d\leq4$.

\subsubsection{Instability of the Ising fixed point and modified Harris criterion}
\label{sec:harris}
The exact relation~\eqref{eq:Zf_exact} allows us to write the $\beta$-function for $\tif_R$ in closed form. Using $Z_f=Z_\tau^2$ one finds
\begin{equation}
    \beta_{\tif} = \tif_R\!\left(\epsilon - 2\gamma_\tau
        + (X_R-1)\gamma_\lambda\right),
    \label{eq:betaftil_exact}
\end{equation}
exact to all orders. At a fixed point, $\gamma_\tau^* = 2 - \nu^{-1}$ and $\gamma_\lambda^* = z-2$, so
\begin{equation}
    \beta_{\tif}^* = \nu^{-1}\tif_R^*\!\left[2 - d\nu
        + \nu(X_R^*-1)(z-2)\right].
    \label{eq:betaf_FP}
\end{equation}

The stability of a $\tif_R^*=0$ fixed point with respect to the density coupling is determined by the sign of the bracket in Eq.~\eqref{eq:betaf_FP}. For fixed points with $X_R^*\neq0$, the term $(X_R^*-1)(z-2)$ vanishes by Eq.~\eqref{eq:betaX_FP}, and the stability condition reduces to $2-d\nu<0$. This is the Harris criterion~\cite{harris1974random}: for the Ising universality class, $\nu d>2$ requires $\nu>2/d$, which fails in $d=3$ where $\nu\approx0.63<2/3$~\cite{pelissetto2002critical}. The Ising fixed point is therefore unstable with respect to the BIM coupling $\tif_R$ in $d=3$.

For fixed points with $X_R^*=0$, the stability condition becomes
\begin{equation}
    2-\nu(d+z-2)<0
    \label{eq:modHarris}
\end{equation}
which is a modified Harris criterion accounting for the dynamic exponent~\cite{vojta2016spatiotemporal}. For the Model A fixed point in $d=3$, using $z\approx2.02$~\cite{hasenbusch2020dynamic,adzhemyan2022dynamic} and $\nu\approx0.63$~\cite{pelissetto2002critical}, one finds $2-\nu(d+z-2)\approx0.10>0$, so Model A is likewise unstable to the coupling $\tif_R$ in $d=3$. 
Applying the same criterion in $d=2$, using the exact value $\nu=1$ and $z\approx 2.17 $~\cite{adzhemyan2022dynamic, Liu2023} for the two-dimensional Ising universality class, gives $2-\nu(d+z-2)\approx-0.17<0$: the density coupling $\tif_R$ is RG-irrelevant at the Ising fixed point in two dimensions, so the BIM does not generate a new universality class in $d=2$. This is consistent with numerical simulations of microscopic BIM realizations~\cite{solon2015flocking,chen2025bim}.

\begin{figure}
\includegraphics[width=\columnwidth]{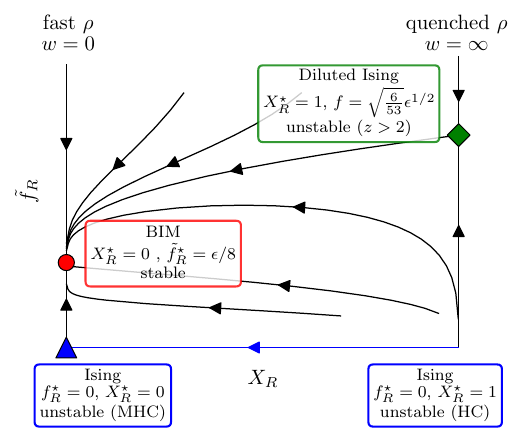}
\caption{{\bf Qualitative sketch of the full renormalization group flow}
On the $\tif_R=0$ manifold (blue line), the Ising universality class is recovered both at $X_R=0$ (fast-diffusion limit) and $X_R=1$ (quenched limit). Both Ising fixed points are however unstable with respect to $\tif_R$: at $X_R=1$ ($w_{R}=\infty$) by the Harris criterion, and at $X_R=0$ by the modified Harris criterion [Eq.~\eqref{eq:modHarris}], since $z>2$. On the quenched manifold $X_{R}=1$, the diluted fixed point (green square, $\tif_R\neq0$) controls the critical dynamics. This fixed point is however unstable with respect to finite defect motility $X_{R}\neq1$, since $z>2$ at the quenched fixed point. The BIM fixed point (red circle) is the only stable fixed point in both the $X_R$ and $\tif_R$ directions, and controls the critical behavior of the Brownian Ising model in $d<4$.}
\label{fig:mockFlow}
\end{figure}

Figure~\ref{fig:mockFlow} summarizes the schematic structure of the RG flow in the $(X_R,\tif_R)$ plane, inferred from the stability criteria and the two-loop calculation. The crossover between the two fixed points cannot be captured within the standard Callan-Symanzik framework, and requires a double expansion in both $\epsilon$ and $1/w_R$, which is beyond the scope of this work \cite{FREY199552, cavagna2019renormalization}.

\subsection{Exact exponent relation}
\label{sec:exactnu}
For any fixed point with $X_R^*<1$, the symmetry~\eqref{eq:shift_sym} holds, and Eq.~\eqref{eq:betaf_FP} is valid at the fixed point with finite $\tif_R^*$. Setting $\beta_{\tif}^*=0$ with $\tif_R^*\neq0$ gives the exact relation
\begin{equation}
    \nu = \frac{2}{d + z - 2}\,,
    \label{eq:nu_exact}
\end{equation}
valid to all orders in perturbation theory at the BIM fixed point. This relation connects the static exponent $\nu$ to the dynamic exponent $z$: it is satisfied by the one-loop results ($\nu=\frac{1}{2}+\frac{\epsilon}{8}$, $z=2$) and allows the two-loop correction to $\nu$ to be read off directly from Eq.~\eqref{eq:z_2loop}:
\begin{equation}
    \nu = \frac{1}{2} + \frac{\epsilon}{8}
        + \frac{\epsilon^2}{768}\!\left(29 - 24\log\!\tfrac{4}{3}\right)
        + \mathcal{O}(\epsilon^3)\,.
    \label{eq:nu_2loop}
\end{equation}
Evaluating at $d=3$ ($\epsilon=1$) yields
\begin{align}
    \nu \approx 0.653\,, \qquad z \approx 2.019\,.
    \label{eq:BIM_d3}
\end{align}
\section{Discussion and Conclusions}
\label{sec:conclusion}

We have carried out a systematic field-theoretical renormalization-group analysis of the Brownian Ising model (BIM), a paradigmatic system in which a diffusing scalar density field couples to an Ising order parameter without satisfying detailed balance.
The coupling is structurally simple: the local density $\rho$ shifts the effective mass of the order parameter $\psi$ and diffuses independently, without being driven by $\psi$.  This one-sided coupling suffices to break the fluctuation-dissipation theorem (FDT) and to generate a new universality class.

\subsection*{Summary of results}

The RG analysis, carried out within the $\epsilon=4-d$ expansion, reveals four fixed points (Table~\ref{tab:FP}): the Gaussian, Model~A, and quenched-disorder (diluted Ising) fixed points, all familiar, together with a fourth that is genuinely new and entirely due to the non-equilibrium density coupling. At one loop the BIM fixed point sits at $X_R^*=0$, $u_R^*=\epsilon/6$, $\tif_R^*=\epsilon/8$, with exponents $\nu=\tfrac12+\epsilon/8$, $\eta=-\epsilon/8$, and $\eta'\equiv 2-\gamma/\nu = 0$, $z=2$. Its most striking feature is the inequality $\eta\neq\eta'$: correlation and response functions acquire different anomalous dimensions, a direct signature of FDT violation that cannot occur in any equilibrium model.

Because $z=2$ and $\eta'=0$ at one loop, the leading corrections to both first arise at two loops, and follow from just two diagrams for $\Gamma_{\psi\hpsi}$ at $X_R=0$ — causality eliminating all other contributions (Appendix~\ref{app:twoloops}). The resulting $\mathcal{O}(\epsilon^2)$ corrections to $z$ and $\eta'$ [Eqs.~\eqref{eq:z_2loop}--\eqref{eq:etap_2loop}; Table~\ref{tab:exp}] are both positive, so $z>2$. The same correction lifts the one-loop marginal direction, giving a positive correction-to-scaling exponent $\omega$ and confirming the BIM fixed point as the unique, stable IR attractor at order $\epsilon^2$. In $d=3$ ($\epsilon=1$) the exponents evaluate to $z\approx2.019$, $\nu\approx0.653$, $\eta\approx-0.125$, and $\omega\approx0.020$; the smallness of $\omega$ implies a very slow approach to asymptotic scaling, with consequences for numerical tests discussed below.

\subsection*{Exact relations and their consequences}

A central contribution of this work is a set of relations among renormalization factors that hold exactly, to all orders in perturbation theory. The non-renormalization of the density sector gives $Z_{\hrho}=Z_\rho^{-1}$, $Z_w=Z_\lambda$, and $Z_f=Z_\rho$ [Eq.~\eqref{eq:Zw_exact}], while the emergent shift symmetry $(\rho,\tau)\to(\rho+\phi,\tau-\phi)$ implies $Z_f=Z_\tau^2$ [Eqs.~\eqref{eq:shift_sym},~\eqref{eq:Zf_exact}]. Together they fix the exact $\beta$-function $\beta_X= X_R(X_R-1)\gamma_\lambda$ [Eq.~\eqref{eq:betaX_exact}], which sorts any stable fixed point into one of three classes --- $X_R^*=0$, $X_R^*\in(0,1)$, or $X_R^*=1$, with $z>2$, $z=2$, or $z<2$ respectively --- and a closed-form exact $\beta$-function for $\tif_R$ [Eq.~\eqref{eq:betaftil_exact}].

These relations let us reach conclusions valid well beyond the small-$\epsilon$ limit. The exact $\beta_{\tif}$ yields a Harris-type stability criterion: the equilibrium Ising fixed point requires $\nu(d+z-2)>2$ to be stable [Eq.~\eqref{eq:modHarris}], which fails in $d=3$. The exact form of $\beta_X$ rules out the quenched (diluted-Ising) fixed point, which has $X_R^*=1$ but $z>2$ in all $d\leq4$: it is unstable to any finite motility, and the quenched limit serves only as a consistency check, reproducing the degenerate one-loop equations of the diluted Ising model [Eqs.~\eqref{eq:betau_quenched}--\eqref{eq:betaf_quenched}]. With the three competing fixed points excluded, the BIM fixed point is established as the unique IR attractor for any non-zero diffusion constant. Finally, imposing $\beta_{\tif}^*=0$ at finite $\tif_R^*$ gives the exact relation $\nu=2/(d+z-2)$ [Eq.~\eqref{eq:nu_exact}] between the static and dynamic exponents, which reduces the number of independent exponents and fixes the two-loop value of $\nu$ from the two-loop value of $z$ at no extra cost.

\subsection*{Outlook}
Several open questions remain.
First, the perturbative expansion around $d=4$ cannot rule out the existence of additional fixed points at intermediate dimension, e.g., for some $3<d^*<4$. Although this is not often the case (see the Ising and $O(n)$ cases~\cite{pelissetto2002critical}), it cannot be excluded a priori, as the cases of the Random-Field Ising model \cite{parisi1979random,bray1985scaling} and the Branching and Annihilating Random walks~\cite{cardy1996branching} demonstrate. Non-perturbative renormalization-group methods (NPRG) would be required to address this question.

Second, the small value of $\omega\approx0.020$ in $d=3$ implies very slow crossover from the mean-field regime to the true asymptotic scaling, which may render the BIM universality class difficult to detect in numerical or experimental studies with finite systems. Higher-order calculations or resummation techniques could sharpen the prediction for $\omega$.

Third, our analysis is restricted to the case of purely passive density fluctuations, where the density evolves independently of the order parameter.  More general couplings, including active feedback from $\psi$ onto $\rho$ (as in Model~C), could give rise to further universality classes and will be explored in future work.

More broadly, the BIM provides a conceptually clean example of how a conserved diffusive mode can destabilize equilibrium criticality, even when its relaxation is fast. The mechanism — asymmetric coupling between a non-conserved order parameter and a conserved density — is generic and should operate in a wide class of driven, motile, or environmentally fluctuating systems.  We hope that the exact results and two-loop exponents derived here will serve as benchmarks for future numerical, experimental, and theoretical investigations of non-equilibrium critical phenomena.

\section*{Acknowledgment}
We thank Michael E. Cates and Johannes Pausch for useful discussion and guidance in initiating this project. We also thank Andrea Cavagna, Tomas S. Grigera, Q. Yu for useful discussions. MS thanks Rosalba Garcia-Millan for discussions in the early stages of this work. MS is grateful to the Department of Applied Mathematics and Theoretical Physics at the University of Cambridge, where the early stage of this work was performed, for their support. MS acknowledges support from ERC Advanced Grant RG.BIO (no. 785932). LDC was supported by the National Science Foundation, through the Center for the Physics of Biological Function (PHY--1734030).

\appendix

\section{Diagrammatic perturbation theory}
\subsection{Feynman rules}\label{app:feynmanrules}
Diagrammatically, the bare propagators and correlators in Eqs.~\eqref{eq:Gpsi}--\eqref{eq:Crho0} will be represented by lines, joining the two respective fields:
\begin{equation}
    G_\psi^0\, :\,
    \begin{tikzpicture}[baseline=-\the\dimexpr\fontdimen22\textfont2\relax]
        \begin{feynman}[baseline=(v1.base)]
            \vertex (v1) at (-1,0) {$\hpsi (\tk)$};
            \vertex (v2) at (1,0) {$\psi (\tq)$};
            \diagram*{
                (v1) -- [fermion] (v2),
            };
        \end{feynman}
    \end{tikzpicture}
    =\langle\hpsi(\tk)\,\psi(\tq)\rangle_0
\end{equation}
\begin{equation}
    C_\psi^0\, :\,
    \begin{tikzpicture}[baseline=-\the\dimexpr\fontdimen22\textfont2\relax]
        \begin{feynman}[baseline=(v1.base)]
            \vertex (v1) at (-1,0) {$\psi (\tk)$};
            \vertex (v2) at (1,0) {$\psi (\tq)$};
            \diagram*{
                (v1) -- (v2),
            };
        \end{feynman}
    \end{tikzpicture}
    =\langle\psi(\tk)\,\psi(\tq)\rangle_0
\end{equation}
\begin{equation}
    G_\rho^0\, :\,
    \begin{tikzpicture}[baseline=-\the\dimexpr\fontdimen22\textfont2\relax]
        \begin{feynman}[baseline=(v1.base)]
            \vertex (v1) at (-1,0) {$\hrho (\tk)$};
            \vertex (v2) at (1,0) {$\rho (\tq)$};
            \diagram*{
                (v1) -- [charged boson] (v2),
            };
        \end{feynman}
    \end{tikzpicture}
    =\langle\hrho(\tk)\,\rho(\tq)\rangle_0
\end{equation}
\begin{equation}
    C_\rho^0\, :\,
    \begin{tikzpicture}[baseline=-\the\dimexpr\fontdimen22\textfont2\relax]
        \begin{feynman}[baseline=(v1.base)]
            \vertex (v1) at (-1,0) {$\rho (\tk)$};
            \vertex (v2) at (1,0) {$\rho (\tq)$};
            \diagram*{
                (v1) -- [boson] (v2),
            };
        \end{feynman}
    \end{tikzpicture}
    =\langle\rho(\tk)\,\rho(\tq)\rangle_0
\end{equation}
The arrow convention follows from the direction of time in the real-space representation: the retarded propagators $G_\psi^0$ and $G_\rho^0$ vanish by causality when $\hpsi$ and $\hrho$ are evaluated at later times than $\psi$ and $\rho$, respectively.

The vertices corresponding to the interaction terms in $\mathcal{L}_I'$ will be represented by points where lines meet:
\begin{equation}
    \begin{tikzpicture}[baseline=-\the\dimexpr\fontdimen22\textfont2\relax]
        \begin{feynman}
            \vertex (v1) at (-0.73,1) {$\psi (\tk)$};
            \vertex (v2) at (-0.73,-1) {$\rho (\tp)$};
            \vertex (v3) at (1,0) {$\hpsi (\tq)$};
            \vertex [empty dot] (i) at (0,0) {};
            \diagram*{
                (v1) -- (i) -- [boson] (v2),
                (i) -- [fermion] (v3),
            };
        \end{feynman}
    \end{tikzpicture}
    =-\tilde\delta(\tk + \tp + \tq)
    \label{eq:vertex_f}
\end{equation}
\begin{equation}
    \begin{tikzpicture}[baseline=-\the\dimexpr\fontdimen22\textfont2\relax]
        \begin{feynman}
            \vertex (v1) at (-1,1.4) {$\psi (\tk)$};
            \vertex (v2) at (-1.3,0) {$\psi (\tp)$};
            \vertex (v3) at (-1,-1.4) {$\psi (\thh)$};
            \vertex (v4) at (1.3,0) {$\hpsi (\tq)$};
            \vertex [empty dot] (i) at (0,0) {};
            \diagram*{
                (v1) -- (i) -- (v2),
                (v3) -- (i),
                (i) -- [fermion] (v4),
            };
        \end{feynman}
    \end{tikzpicture}
    =-u_{0}\tilde\delta(\tk + \tp + \tq + \thh)
    \label{eq:vertex_u}
\end{equation}
The negative signs in the vertex reflect the fact that the action is $S = \dots + \int\!\di^d x\,\di t\,\mathcal{L}_{I}^\prime$ and the path integral weight is $\eu^{-S}$. A negative sign therefore naturally arises in the perturbative expansion of the path integral, and will be included in the definition of the vertices. The delta functions enforce momentum and frequency conservation at the vertices, reflecting locality in space and time of the action.

Each term in the perturbative expansion can be diagrammatically expressed as a set of $n$ vertices joined by propagators or correlators, with each closed loop corresponding to an integration over internal momenta and frequencies. The combinatorial factors associated with each diagram are determined by the symmetry factors of the diagram and the number of ways it can be generated from the interaction terms. These diagrammatic rules will be used to systematically compute perturbative corrections to the vertex functions $\Gamma$.

\subsection{One-loop divergent diagrams}\label{app:oneloop}
Here we list the diagrammatic contributions up to one loop to the diverging vertex-functions. We then also provide their divergent parts, expressed as poles in $\epsilon=4-d$, and how they are reabsorbed in the renormalization factors.

\subsubsection{Diagrammatic expansion of vertex functions}
The diagrams contributing to $\Gamma_{\psi\hpsi}$ are:
\begin{equation*}
\begin{split}
    \Gamma_{\psi\hpsi}=&
    -\bigl(
    \begin{tikzpicture}[baseline=-\the\dimexpr\fontdimen22\textfont2\relax]
        \begin{feynman}[baseline=(v1.base)]
            \vertex (v1) at (0,0);
            \vertex (vi) at (1,0);
            \diagram* {
                (v1) -- [fermion] (vi) 
            };
        \end{feynman}
    \end{tikzpicture}
    \bigr)^{-1}
    +
    \begin{tikzpicture}[baseline=-\the\dimexpr\fontdimen22\textfont2\relax]
        \begin{feynman}[baseline=(v1.base)]
            \vertex (i1) at (-0.4,0);
            \vertex (i2) at (0.6,0);
            \vertex (v1) at (0,0);
            \vertex (vi) at (0,1);
            \diagram* {
                (i1) -- (v1),
                (v1) -- [half left,out=45,in=90] (vi),
                (vi) -- [half left,out=90,in=135] (v1),
                (v1) -- [fermion] (i2)
            };
        \end{feynman}
    \end{tikzpicture}
    +
    \begin{tikzpicture}[baseline=-\the\dimexpr\fontdimen22\textfont2\relax]
        \begin{feynman}[baseline=(v1.base)]
            \vertex (i1) at (-0.4,0);
            \vertex (i2) at (1.8,0);
            \vertex (v1) at (0,0);
            \vertex (viup) at (0.75,0.5);
            \vertex (vidown) at (0.75,-0.5);
            \vertex (v2) at (1.2,0);
            \diagram* {
                (i1) -- (v1),
                (v1) -- [fermion, half left] (v2),
                (viup) -- [out=0,in=90] (v2),
                (v1) -- [photon, half right] (v2),
                (v2) -- [fermion] (i2)
            };
        \end{feynman}
    \end{tikzpicture}
\end{split}
\end{equation*}
Which leads to
\begin{equation}
    \begin{split}
     &\Gamma_{\psi\hpsi}(q,\omega)=\iu \lambda_0^{-1}\omega-q^2+\tau_0 -  
    3 u_0 \int d^dp\frac1{p^2+\tau_0} \\
     &+ f_0 \int d^dp \frac{w_0^2}{p^2+w_0 \left((p+q)^2+\tau_0 -\iu \lambda_0^{-1}\omega \right)}
\end{split}
\end{equation}
Three renormalization factors are required to ensure the finiteness of $\Gamma_{\psi\hpsi}$, as only $\Gamma_{\psi\hpsi}(0,0)$, $\partial_\omega \Gamma_{\psi\hpsi}(0,0)$ and $\partial_{q^2} \Gamma_{\psi\hpsi}(0,0)$ are divergent.

The diagrams contributing to $\Gamma_{\hpsi\hpsi}$ are:
\begin{equation*}
\begin{split}
    \Gamma_{\hpsi \hpsi}&=
    \begin{tikzpicture}[baseline=-\the\dimexpr\fontdimen22\textfont2\relax]
        \begin{feynman}[baseline=(v1.base)]
            \vertex (v1) at (0,0);
            \vertex (vi) [dot] at (1,0) {};
            \vertex (v2) at (2,0);
            \diagram* {
                (v1) -- [anti fermion] (vi),
                (v2) -- [anti fermion] (vi)
            };
        \end{feynman}
    \end{tikzpicture}
    +
    \begin{tikzpicture}[baseline=-\the\dimexpr\fontdimen22\textfont2\relax]
        \begin{feynman}[baseline=(v1.base)]
            \vertex (i1) at (-0.75,0);
            \vertex (i2) at (2.25,0);
            \vertex (v1) at (0,0);
            \vertex (viup) at (0.75,0.5);
            \vertex (vidown) at (0.75,-0.5);
            \vertex (v2) at (1.5,0);
            \diagram* {
                (i1) -- [anti fermion] (v1),
                (v1) -- [out=90,in=180] (viup),
                (viup) -- [out=0,in=90] (v2),
                (v1) -- [photon,out=270,in=180] (vidown),
                (vidown) -- [photon,out=0,in=270] (v2),
                (i2) -- [anti fermion] (v2)
            };
        \end{feynman}
    \end{tikzpicture}
\end{split}
\end{equation*}
This vertex function is divergent only at $q=\omega=0$, so it is sufficient to look at the vertex function at zero external frequency and momenta, which leads to
\begin{equation}
    \begin{split}
        \Gamma_{\hpsi \hpsi}&(0,0)=\lambda_0^{-1}+\\
        & + \lambda_0^{-1}f_0\int d^dp \frac{w_0}{\left(p^2+\tau_0 \right) \left(w_0 \left(p^2+\tau_0 \right)+p^2\right)}
    \end{split}
\end{equation}

The diagrams contributing to $\Gamma_{\rho\psi\hpsi}$ are:
\begin{equation*}
\begin{split}
    \Gamma_{\rho\psi\hpsi}&=
    \begin{tikzpicture}[baseline=-\the\dimexpr\fontdimen22\textfont2\relax]
        \begin{feynman}[baseline=(v1.base)]
            \vertex (i1) at (-0.8,0);
            \vertex (i2) at (0,1);
            \vertex (v1) at (0,0);
            \vertex (i3) at (0.8,0);
            \diagram* {
                (i1) -- (v1),
                (i2) -- [boson] (v1),
                (v1) -- [fermion] (i3)
            };
        \end{feynman}
    \end{tikzpicture}
    +
    \begin{tikzpicture}[baseline=-\the\dimexpr\fontdimen22\textfont2\relax]
        \begin{feynman}[baseline=(v1.base)]
            \vertex (i1) at (-0.8,0.0);
            \vertex (i2) at (1,0);
            \vertex (i3) at (0,2.25);
            \vertex (v1) at (0,0);
            \vertex (viup) at (-0.5,0.75);
            \vertex (vidown) at (0.5,0.75);
            \vertex (v2) at (0,1.5);
            \diagram* {
                (i1) -- (v1) -- [fermion] (i2),
                (v1) -- [out=180,in=270] (viup) -- [out=90,in=180] (v2) -- [fermion,out=0,in=90] (vidown) -- [out=270,in=0] (v1),
                (v2) --[photon] (i3),
            };
        \end{feynman}
    \end{tikzpicture}
    +
    \begin{tikzpicture}[baseline=-\the\dimexpr\fontdimen22\textfont2\relax]
        \begin{feynman}[baseline=(v1.base)]
            \vertex (i1) at (0,2);
            \vertex (i2) at (1.5,0);
            \vertex (i3) at (-1.2,0);
            \vertex (v1) at (0,1.3);
            \vertex (vi) at (0.375,0.65);
            \vertex (vj) at (-0.375,0.65);
            \vertex (v3) at (-0.75,0);
            \vertex (v2) at (0.75,0);
            \diagram* {
                (i1) -- [photon] (v1),
                (v2) -- [fermion] (i2),
                (i3) -- [] (v3),
                (v1) -- [fermion] (vi) -- (v2) -- [photon] (v3) -- [fermion] (vj) -- (v1)
            };
        \end{feynman}
    \end{tikzpicture}
\end{split}
\end{equation*}
In $\Gamma_{\rho\psi\hpsi}$ the divergence appears at zero external momenta and frequencies, hence
\begin{equation}
    \begin{split}
        \Gamma_{\rho\psi\hpsi}(0,0,0,0)&=-1 + 3 u_0\int d^dp \frac{1}{\left(p^2+\tau_0 \right)^2}-\\
        &-f_0\int d^dp \frac{w_0^2}{\left(w_0 (p^2 +\tau_0) + p^2\right)^2}
    \end{split}
\end{equation}

The diagrams contributing to $\Gamma_{\psi\psi\psi\hpsi}$ are:
\begin{equation*}
\begin{split}
    \Gamma_{\psi\psi\psi\hpsi}&=
    \begin{tikzpicture}[baseline=-\the\dimexpr\fontdimen22\textfont2\relax]
        \begin{feynman}[baseline=(v1.base)]
            \vertex (i1) at (-0.75,0.75);
            \vertex (i2) at (-0.75,-0.75);
            \vertex (v1) at (0,0);
            \vertex (i3) at (0.75,0.75);
            \vertex (i4) at (0.75,-0.75);
            \diagram* {
                (i1) -- (v1),
                (i2) -- (v1),
                (v1) -- [fermion] (i3),
                (v1) -- (i4)
            };
        \end{feynman}
    \end{tikzpicture}
    +
    \begin{tikzpicture}[baseline=-\the\dimexpr\fontdimen22\textfont2\relax]
        \begin{feynman}[baseline=(v1.base)]
            \vertex (i1) at (-0.75,0.5);
            \vertex (i2) at (-0.75,-0.5);
            \vertex (i3) at (2.25,0.5);
            \vertex (i4) at (2.25,-0.5);
            \vertex (v1) at (0,0);
            \vertex (viup) at (0.75,0.5);
            \vertex (vidown) at (0.75,-0.5);
            \vertex (v2) at (1.5,0);
            \diagram* {
                (i1) -- (v1),
                (i2) -- [] (v1),
                (v1) -- [fermion,out=90,in=180] (viup),
                (viup) -- [out=0,in=90] (v2),
                (v1) -- [out=270,in=180] (vidown),
                (vidown) -- [out=0,in=270] (v2),
                (v2) --[fermion]  (i3),
                (v2) --[] (i4)
            };
        \end{feynman}
    \end{tikzpicture}
    + \\ 
    &+ 
    \begin{tikzpicture}[baseline=-\the\dimexpr\fontdimen22\textfont2\relax]
        \begin{feynman}[baseline=(v1.base)]
            \vertex (i1) at (-0.75,0.5);
            \vertex (i2) at (-0.75,-0.5);
            \vertex (i3) at (1.5,1);
            \vertex (i4) at (1.5,-1);
            \vertex (v1) at (0,0);
            \vertex (viup) at (0.5,0.3);
            \vertex (vidown) at (0.5,-0.3);
            \vertex (v2) at (1,0.6);
            \vertex (v3) at (1,-0.6);
            \diagram* {
                (i1) -- (v1),
                (i2) -- [] (v1),
                (v1) -- [fermion] (viup) --  (v2),
                (v1) -- [] (vidown) -- [anti fermion] (v3),
                (v2) -- [photon] (v3),
                (v2) -- [fermion] (i3),
                (v3) --[] (i4)
            };
        \end{feynman}
    \end{tikzpicture}
    +
    \begin{tikzpicture}[baseline=-\the\dimexpr\fontdimen22\textfont2\relax]
        \begin{feynman}[baseline=(v1.base)]
            \vertex (i1) at (-0.5,1);
            \vertex (i2) at (1.75,0.5);
            \vertex (i3) at (-0.5,-1);
            \vertex (i4) at (1.75,-0.5);
            \vertex (v1) at (0,0.6);
            \vertex (vi) at (0.5,0.3);
            \vertex (vj) at (0.5,-0.3);
            \vertex (v3) at (0,-0.6);
            \vertex (v2) at (1,0);
            \diagram* {
                (i1) -- (v1),
                (v2) -- [fermion] (i2),
                (i3) -- [] (v3),
                (i4) --  (v2),
                (v1) -- [fermion] (vi) --  (v2),
                (v3) -- [fermion] (vj) -- (v2),
                (v3) -- [photon] (v1)
            };
        \end{feynman}
    \end{tikzpicture}
\end{split}
\end{equation*}
In $\Gamma_{\psi\psi\psi\hpsi}$ as well the divergence appears at zero external momenta and frequencies
\begin{widetext}
\begin{equation}
\begin{split}
    \Gamma_{\psi\psi\psi\hpsi}(0,0,0,0,0,0)=-u_0 + &9 u_0^2\int d^dp \frac{1}{\left(p^2+\tau_0 \right)^2} -3 u_0 f_0\int d^dp \frac{w_0}{\left(p^2+\tau_0 \right) \left(w_0 (p^2+\tau_0)+p^2\right)}-\\
    -&3 u_0 f_0\int d^dp \frac{w_0^2}{\left(w_0 (p^2+\tau_0)+p^2\right)^2}
\end{split}
\end{equation}
\end{widetext}

\subsubsection{$Z$ factors}
Divergences in the diagrams can be removed by multiplicative renormalization, introducing the following $Z$-factors
\begin{equation}
    Z_{\hpsi}^{1/2}Z_{\psi}^{1/2}Z_\lambda^{-1}=1+\frac{f_0 w_0^2}{(w_0+1)^2}\frac{K_d\mu ^{-\epsilon }}{\epsilon}
\end{equation}
\begin{equation}
    Z_{\hpsi}^{1/2}Z_{\psi}^{1/2}=1+\frac{f_0 w_0^2}{(w_0+1)^3}\frac{K_d\mu ^{-\epsilon }}{\epsilon}
\end{equation}
\begin{equation}
    Z_{\hpsi}^{1/2}Z_{\psi}^{1/2}Z_\tau=1-3u_0\frac{K_d\mu ^{-\epsilon }}{\epsilon}+\frac{f_0 w_0^2}{(w_0+1)^2}\frac{K_d\mu ^{-\epsilon }}{\epsilon}
\end{equation}
\begin{equation}
    Z_{\hpsi}Z_\lambda^{-1}=1+\frac{f_0 w_0}{w_0+1}\frac{K_d\mu ^{-\epsilon }}{\epsilon}
\end{equation}
\begin{equation}
\begin{split}
    Z_{\hpsi}^{1/2}Z_{\psi}^{3/2}Z_u&=1+\frac{3 f_0 w_0^2}{(w_0+1)^2} \frac{K_d\mu ^{-\epsilon }}{\epsilon} +\\
    &+\frac{3 f_0 w_0}{(w_0+1)} \frac{K_d\mu ^{-\epsilon }}{\epsilon} - 9 u_0 \frac{K_d\mu ^{-\epsilon }}{\epsilon}
\end{split}
\end{equation}
\begin{equation}
    Z_{\hpsi}^{1/2}Z_{\psi}^{1/2}Z_\rho^{1/2}=1+\frac{f_0 w_0^2}{(w_0+1)^2}\frac{K_d\mu ^{-\epsilon }}{\epsilon}-3 u_0\frac{K_d\mu ^{-\epsilon }}{\epsilon}
\end{equation}
\begin{equation}
    Z_{\hrho}^{1/2}Z_\rho^{1/2}Z_\lambda^{-1}Z_w=1
    \label{eq:omegarho}
\end{equation}
\begin{equation}
    Z_{\hrho}^{1/2}Z_\rho^{1/2}=1
\end{equation}
\begin{equation}
    Z_{\hrho}Z_\lambda^{-1}Z_wZ_f=1
    \label{eq:noiserho}
\end{equation}

\subsection{Two-loop}\label{app:twoloops}
The two-loop calculation is performed to assess the stability of the $w_{R}^*=0$ fixed point. We shall therefore perform the calculation directly at $w_0=0$, but with $\tif_0=f_0w_0/(1+w_0)$ (see Sec.~\ref{sec:betaBIM}). In this limit, $\rho$ becomes a random noise with delta-correlations in time. Hence, any $\rho$ correlation function
\begin{equation}
    C_{\rho}:\quad
    \begin{tikzpicture}[baseline=(a.base)]
        \begin{feynman}
            \vertex (a) at (0,0);
            \vertex (b) at (1,0);
            \diagram* {
                (a) -- [boson] (b),
            };
        \end{feynman}
    \end{tikzpicture}
\end{equation}
forces the two connected points to have the same time. If this causes a closed loop of many propagators of $\psi$, the diagram vanishes because of causality. An example of such a vanishing diagram is
\begin{equation}
    \begin{tikzpicture}[baseline=-\the\dimexpr\fontdimen22\textfont2\relax]
        \begin{feynman}[baseline=(v1.base)]
            \vertex (i1) at (-1,0);
            \vertex (i2) at (3,0);
            \vertex (v1) at (0,0);
            \vertex (v2) at (2,0);
            \vertex (viup) at (1,0.75);
            \vertex (vidown) at (1, -1);
            \diagram* {
                (i1) -- (v1),
		(v1) -- [anti fermion, red, quarter right] (vidown),
		(vidown) -- [ photon, quarter right] (v2),
		(v2) -- [anti fermion, red, half right] (v1), 
		(v2) --[fermion] (i2) ,
		(v1) -- [quarter left](vidown)
            };
        \end{feynman}
    \end{tikzpicture}
\end{equation}
We have highlighted in red the closed loop of propagators that vanishes when the correlations of $\rho$ becomes proportional to $\delta(t-t')$.

Assessing the stability of the BIM fixed point requires only the evaluation of two-loop diagrams of $\Gamma_{\psi\hpsi}$. In particular, only the second-order contributions to $\partial_\omega\Gamma_{\psi\hpsi}(0,0)$ and $\partial_{q^2}\Gamma_{\psi\hpsi}(0,0)$ are needed to determine the first non-trivial contributions to $z$ (and hence $\omega$) and $\eta'$.
With this in mind, there is only one additional diagram compared to the usual computation of Model A, and the vertex function $\Gamma_{\psi\hpsi}$, at two loops, is thus given by
\begin{equation}
    \Gamma_{\psi \hpsi} = \cdots +
    \begin{tikzpicture}[baseline=-\the\dimexpr\fontdimen22\textfont2\relax]
        \begin{feynman}[baseline=(v1.base)]
            \vertex (i1) at (-0.5,0);
            \vertex (i2) at (2.2,0);
            \vertex (v1) at (0,0);
            \vertex (v2) at (1.5,0);
            \diagram* {
                (i1) -- (v1),
		(v1) -- [fermion, half left] (v2),
		(v1) -- (v2) ,
		(v2) --[fermion] (i2) ,
		(v1) -- [half right](v2)
            };
        \end{feynman}
    \end{tikzpicture}
 + 
    \begin{tikzpicture}[baseline=-\the\dimexpr\fontdimen22\textfont2\relax]
        \begin{feynman}[baseline=(v1.base)]
            \vertex (i1) at (-0.5,0);
            \vertex (i2) at (2,0);
            \vertex (v1) at (0,0);
            \vertex (v2) at (1.5,0);
            \vertex (viup) at (0.75,0.5625);
            \vertex (vidown) at (0.75, -0.75);
            \diagram* {
                (i1) -- (v1),
		(v1) -- [ boson, quarter right] (vidown),
		(vidown) -- [ fermion, quarter right] (v2),
		(v1) -- [fermion, half left] (v2), 
		(v2) --[fermion] (i2) ,
		(v2) -- [quarter right](vidown)
            };
        \end{feynman}
    \end{tikzpicture}
\end{equation}
where the ellipsis stands for all contributions up to one-loop. The calculation of these diagrams leads to
\begin{widetext}
\begin{equation}
    \begin{split}
	\Gamma_{\psi\hpsi}(q,\omega) ^{(2)}= & 18 u_0^2 \int d^dp d^dk \frac{1}{\left(k^2+\tau_0 \right) \left(p^2+\tau_0 \right) \left(p^2+k^2+(q-k-p)^2+3 \tau_0 -\iu\lambda_0^{-1} \omega \right)} +\\
        &-12 u_0\tif_0\int d^dp d^dk\frac{1}{q^2 \left(p^2+\tau_0 \right) \left((k-q)^2+p^2+(q-p)^2+3 \tau_0 -\iu\lambda_0^{-1} \omega \right)}
    \end{split}
\end{equation}
\end{widetext}
Here, the superscript $(2)$ indicates that we are only considering the two-loop contributions. The first term is the usual two-loop contribution of Model A, while the second term is the new contribution due to the coupling with $\rho$.

Direct evaluation of these diagrams leads to the following contributions
\begin{equation}
    \begin{split}
	Z_{\hpsi}^{1/2}Z_{\psi}^{1/2}Z_\lambda^{-1}=1&+\frac{9}{2}\log \left(\frac{4}{3}\right)\frac{ u_0^2 K_d^2  \mu ^{-2 \epsilon }}{ \epsilon }-
    \\ & -3\log \left(\frac{4}{3}\right)\frac{\tif_0 u_0  K_d^2\mu ^{-2 \epsilon }}{\epsilon}
    \end{split}
\end{equation}
\begin{equation}
    \begin{split}
	Z_{\hpsi}^{1/2}Z_{\psi}^{1/2}=1 & +3\left(\frac{1}{12}-\log \left(\frac{4}{3}\right)\right) \frac{\tif_0  u_0  K_d^2 \mu ^{-2 \epsilon }}{\epsilon }+ \\
    & +\frac{3}{4}\frac{ u_0^2 K_d^2 \mu ^{-2 \epsilon }}{\epsilon }
    \end{split}
\end{equation}

\section{Diagrams in the quenched limit}\label{app:quenched}
The new vertex of the quenched action, Eq.~\eqref{eq:SI_quenched}, is diagrammatically represented by two vertices~\eqref{eq:vertex_f} connected by a density propagator $C_\rho$. It reads
\begin{equation}
    \begin{tikzpicture}[baseline=-\the\dimexpr\fontdimen22\textfont2\relax]
        \begin{feynman}[baseline=(v1.base)]
            \vertex (i1) at (1,0.75) {$\hpsi(\tq)$};
            \vertex (i2) at (-1,0.75) {$\psi (\tk)$};
            \vertex (v1) at (0,0.75);
            \vertex (v2) at (0,-0.75);
            \vertex (i3) at (1,-0.75) {$\hpsi (\tp)$};
            \vertex (i4) at (-1,-0.75) {$\psi (\thh)$};
            \diagram* {
                (v1) -- [fermion] (i1),
                (i2) -- (v1),
                (v2) -- [boson] (v1),
                (v2) -- [fermion] (i3),
                (i4) -- (v2),
            };
        \end{feynman}
    \end{tikzpicture}
    =-\frac{C_{\rho}(\tq - \tk)}{2} \tilde\delta(\tq + \tk + \tp + \thh)
\end{equation}
with $C_{\rho}(\tq)=2\pi\tif_0\,\delta(\omega_{q})$ in the quenched limit.

In the formulation given by Eq.~\eqref{eq:S0_quenched}--\eqref{eq:SI_quenched}, the renormalization of the theory is the same as the non-quenched one for $\Gamma_{\psi\hpsi}$ and $\Gamma_{\psi\psi\psi\hpsi }$. The crucial difference is that the divergences of $\Gamma_{\rho \psi\hpsi}$ now appear in the $\Gamma_{\psi\psi\hpsi\hpsi}$ 1PI function.

In the quenched limit, the divergent one-loop diagrams contributing to $\Gamma_{\psi\psi\hpsi\hpsi}$ are
\begin{equation}
\begin{split}
    \Gamma_{\psi\psi\hpsi\hpsi}=
    \begin{tikzpicture}[baseline=-\the\dimexpr\fontdimen22\textfont2\relax]
        \begin{feynman}[baseline=(v1.base)]
            \vertex (i1) at (0.6,0.75);
            \vertex (i2) at (-0.6,0.75);
            \vertex (v1) at (0,0.75);
            \vertex (v2) at (0,-0.75);
            \vertex (i3) at (0.6,-0.75);
            \vertex (i4) at (-0.6,-0.75);
            \diagram* {
                (v1) -- [fermion] (i1),
                (i2) -- (v1),
                (v2) -- [boson] (v1),
                (v2) -- [fermion] (i3),
                (i4) -- (v2),
            };
        \end{feynman}
    \end{tikzpicture}
    &+
    \begin{tikzpicture}[baseline=-\the\dimexpr\fontdimen22\textfont2\relax]
        \begin{feynman}[baseline=(i1.base)]
            \vertex (i1) at (-1,-0.75);
            \vertex (i2) at (1,-0.75);
            \vertex (i3) at (-1,0.75);
            \vertex (i4) at (1,0.75);
            \vertex (v1) at (0,-0.75);
            \vertex (viup) at (-0.4,-0.25);
            \vertex (vidown) at (0.4,-0.25);
            \vertex (v2) at (0,0.25);
            \vertex (v3) at (0,0.75);
            \diagram* {
                (i1) -- (v1) -- [fermion] (i2),
                (v1) -- [out=180,in=270] (viup) -- [out=90,in=180] (v2) -- [fermion,out=0,in=90] (vidown) -- [out=270,in=0] (v1),
                (v2) --[photon] (v3),
                (i3) -- (v3) -- [fermion] (i4)
            };
        \end{feynman}
    \end{tikzpicture}
    + 
    \begin{tikzpicture}[baseline=-\the\dimexpr\fontdimen22\textfont2\relax]
        \begin{feynman}[baseline=(i3.base)]
            \vertex (i1) at (-1.1,-0.75);
            \vertex (i2) at (1.1,-0.75);
            \vertex (i3) at (-1,0.75);
            \vertex (i4) at (1,0.75);
            \vertex (v1) at (-0.6,-0.75);
            \vertex (vj) at (-0.3,-0.23);
            \vertex (v2) at (0.6,-0.75);
            \vertex (vi) at (0.3,-0.23);
            \vertex (v3) at (0,0.29);
            \vertex (v4) at (0,0.75);
            \diagram* {
                (i1) -- (v1) -- [photon] (v2) -- [fermion] (i2),
                (v1) -- [fermion] (vj) -- (v3) -- [fermion] (vi) -- (v2),
                (v3) -- [photon] (v4),
                (i3) -- (v4) -- [fermion] (i4)
            };
        \end{feynman}
    \end{tikzpicture}
    + \\ 
    & \qquad 
    \\ 
    & + 
    \begin{tikzpicture}[baseline=-\the\dimexpr\fontdimen22\textfont2\relax]
        \begin{feynman}[baseline=(v1.base)]
            \vertex (i1) at (-0.5,0.6);
            \vertex (i2) at (1.7,0.6);
            \vertex (i3) at (1.7,-0.6);
            \vertex (i4) at (-0.5,-0.6);
            \vertex (v1) at (0,0.6);
            \vertex (vi) at (0.6,0.6);
            \vertex (v2) at (1.2,0.6);
            \vertex (v3) at (1.2,-0.6);
            \vertex (vj) at (0.6,-0.6);
            \vertex (v4) at (0,-0.6);
            \diagram* {
                (i1) --  (v1) -- [fermion] (vi) -- (v2) -- [fermion] (i2),
                (i4) -- (v4) -- [fermion] (vj) -- (v3) -- [fermion] (i3),
                (v1)  -- [photon] (v4),
                (v2) -- [photon] (v3)
            };
        \end{feynman}
    \end{tikzpicture}
    +
    \begin{tikzpicture}[baseline=-\the\dimexpr\fontdimen22\textfont2\relax]
        \begin{feynman}[baseline=(v1.base)]
            \vertex (i1) at (-0.5,0.6);
            \vertex (i2) at (1.7,0.6);
            \vertex (i3) at (1.7,-0.6);
            \vertex (i4) at (-0.5,-0.6);
            \vertex (v1) at (0,0.6);
            \vertex (vi) at (0.6,0.6);
            \vertex (v2) at (1.2,0.6);
            \vertex (v3) at (1.2,-0.6);
            \vertex (vj) at (0.6,-0.6);
            \vertex (v4) at (0,-0.6);
            \diagram* {
                (i1) --  (v1) -- [fermion] (vi) -- (v2) -- [fermion] (i2),
                (i4) -- (v4) -- [fermion] (vj) -- (v3) -- [fermion] (i3),
                (v1)  -- [photon] (v3),
                (v4) -- [photon] (v2)
            };
        \end{feynman}
    \end{tikzpicture} \\ 
    & \qquad 
\end{split}
\end{equation}
The first three diagrams give the exact same contribution as the three diagrams in $\Gamma_{\rho\psi\hpsi}$, as can be directly inferred from their structure. The last two diagrams, on the other hand, are divergent \emph{only} in the quenched limit $X_{R}=1$ (which is satisfied by taking $w_{0}\to\infty$), hence giving an extra contribution to $Z_{f}$.

The evaluation of the diagrams leads to
\begin{equation}
\begin{split}
     \Gamma_{\psi\psi\hpsi\hpsi}(0,\dots,0)&=\frac{f_0}{2} - 3 f_{0}u_0\int d^dp \frac{1}{\left(p^2+\tau_0 \right)^2}+\\
     &+2f_{0}^{2}\int d^dp \frac{1}{(p^2 +\tau_0)^2}
\end{split}
\end{equation}
Crucially, this has an extra factor $f_{0}^{2}$ compared to the one-loop contribution to $\Gamma_{\rho\psi\hpsi}$ in the $w_{0}\to\infty$ limit.

Finiteness of $\Gamma_{\psi\psi\hpsi\hpsi}$ is ensured by choosing the renormalization factor for $f$ as
\begin{equation}
    Z_{\hpsi} Z_{\psi} Z_f=1+4f_{0}\frac{K_d\mu ^{-\epsilon }}{\epsilon}-6 u_0\frac{K_d\mu ^{-\epsilon }}{\epsilon}
\end{equation}
which leads to the new $\beta$-function for $\tif$ in the quenched limit \eqref{eq:betaf_quenched}.

\bibliography{references.bib}

\end{document}